# A "Good" Regulator May Provide a World Model for Intelligent Systems

Bradly Alicea[1,2,3] Morgan Hough[1,4], Amanda Nelson[1,5], Jesse Parent[1,6]

**ABSTRACT**

One classic idea from the cybernetics literature is the Every Good Regulator Theorem (EGRT). The EGRT provides a means to identify good regulation, or the conditions under which an agent (regulator) can match the dynamical behavior of a system. We reevaluate and recast the EGRT in a modern context to provide insight into how intelligent autonomous learning systems might utilize a compressed global representation (world model). One-to-one mappings between a regulator (*R*) and the corresponding system (*S*) provide a reduced representation that preserves useful variety to match all possible outcomes of a system. The EGRT also extends to second-order cybernetics, where an internal model (*M*) observes the behavior of *S* and supervises a *S*-*R* closed loop mapping. Secondarily, we demonstrate how physical phenomena such as temporal criticality, non-normal denoising, and alternating procedural acquisition can recast behavior as statistical mechanics and yield regulatory relationships. These diverse physical systems challenge the notion of tightly-coupled good regulation when applied to non-uniform and out-of-distribution phenomena. Overall, we aim to recast the EGRT as a potential approach for developing world models that adapt and respond to a wide range of task environments.

## 1. Introduction

Cybernetic regulation requires models that capture key functional elements of a real world system. Yet models cannot simply be a copy of the system. A good representation of regulated room temperature must provide a means to describe the action of regulation with respect to the measurement of room temperature. To regulate room temperature, thermometers are insufficient as they merely report the current temperature state of the room. Thermometers (*T*) must be coupled to a thermostat (*T*') model that returns room temperature to a setpoint. In short, a larger-scale model of reference is needed to achieve global regulation across a variety of previously experienced contexts. A good or complete representation increases the precision of a generic model (Glanville, 1982). Iterative refinement results in a reinforced model of feedback based on an experience of behaving in the world (Jeon, 2022). This mapping between a system and its model's current state allows us to estimate future states of the system, resulting in a **world model** (Bandaru, 2025). World models provide a frame of reference for phenomena not directly experienced by the model. Ha and Schmidthuber (2018) demonstrate this in the realm of a formal world model using a driving simulation where command inputs are regulated by a compressed representation of the world. Forming a closed loop between the system (command inputs) and model (driving simulation and representation) tends to regulate global behavior with respect to travel paths, boundaries, and surfaces. Yet this model must be as complete as possible to truly serve as a world model. Statistical compression with information loss can be detrimental to building a global representation, particularly in over-estimating the influence of non-emergent macrostates in a dynamical system (Hoel, 2017). Indeed, training set ensembles augmented with a feedback mechanism of synthetic data can prevent model collapse (Feng et.al, 2024).

[1] Orthogonal Research and Education Laboratory, Champaign-Urbana, IL. bradly.alicea@outlook.com
[2] OpenWorm Foundation, Boston, MA.
[3] University of Illinois Urbana-Champaign IL.
[4] NeuroTechX, San Francisco, CA.
[5] University of Michigan, Ann Arbor, MI.
[6] University of California, San Diego, CA.

Fortunately, we can describe a global system-to-model mapping in a more formalized manner. This distinguishes the world model from a standard learning model that optimizes for local task gradients. According to Conant and Ashby (1970) and their notion of a good regulator, every good regulator of a system must contain a good approximation of that system. The Every Good Regulator Theorem (EGRT – Conant, 1969; Ashby, 1969; Conant and Ashby, 1970) defines good regulation as the model of a system resulting in good (or sufficient) regulation of that system. Sufficiency is determined by a system (*S*) providing enough information to the regulator (*R*) so that the regulated system remains within a stable range of parameter values. One way to demonstrate the sufficiency of an *S*-*R* relation that can estimate future states is the Daisyworld model (Lovelock, 1983). Daisyworld relies upon closed-loop feedback to maintain a stable state. The simulation is defined by interactions between discrete states representing the world and global system parameters. While the dynamics depend upon an initial condition, critical parameter values allow good regulation resembling dynamic equilibrium rather than a global optimum. This is sometimes used as an example of rein control, where complementary course corrections are made by *R* to maintain a stable trajectory (Boyle et.al, 2011). While disturbances can drive the S away from equilibrium, its stable states are robust with respect to unexpected phenomena (Wood et.al, 2008). Daisyworld is also a process of allostasis: good allostatic regulation allows *R* to be maintained by one of many nearly-equivalent options (Sennesh et.al, 2022). This enables a dynamical systems property called ergodic behavior, where time-averaged behavior is equivalent for all mappings between *R* and *S*. The good regulation of Conant and Ashby (1970) is realized as *R* pushes *S* towards the nearest stable equilibrium point (Cisek and Kalaska, 2010). Complex physiological processes such as neuroplasticity (McEwen and Gianaros, 2011) and cardiovascular stress (Golbidi et.al, 2015) exemplify allostasis: a regulator supports minimizing prediction errors via matching mechanisms (Schulkin and Sterling, 2019). Yet in many cases, allostatic control is not achievable. The ability to maintain a *S*-*R* relation over time is accomplished through closed-loop (recursive) feedback and self-evaluation. Self-evaluation can be realized through a second-order cybernetic relation (Maturana, 1970, Scott, 2004; Jeon, 2020; Cretu, 2020), in which an observer of the *S*-*R* closed-loop relation supervises feedback and continually evaluates the quality of regulation. Observation is specified as the internal model *M* (Francis and Wonham, 1976), which acts to simulate *R*. In this case, *R* implicitly controls *S* through an internal model of an observer (*M*) of *S* (see Appendix B and Figure 1S).

To understand so-called good regulation in the context of modern Artificial Intelligence (AI) requires us to both re-evaluate the historical concept of the EGRT and incorporate more modern insights from control engineering, complexity theory, and cognitive science. We motivate this understanding by formalizing the EGRT, understanding good regulation in terms of prediction and circularity, and focusing on the adaptive behaviors of *R* relative to a complex *S*. Inspiration from complex systems is brought to bear in the form of temporal criticality, non-normal denoising, and alternating procedural acquisition. In this way, we can understand R as an adaptive system driven by statistical regularities. This leads to a discussion of how world models are consistent with good regulation. Finally, we consider how applied good regulation reveals shortcomings of both the EGRT and modern concepts of intelligence.

### 1.1 Formalizing Good Regulation

As a mapping between real world and model, good regulators (GRs) are a one-to-one relationship which requires preservation of a system's empirical structure. What is needed is an active process of self-organization originating from the system's dynamics. Implementing the EGRT also allows us to go beyond this simplistic conception of the world to characterize data inputs in novel ways. Ashby's coupling mechanism (Ashby, 1991) states that self-organization can only occur when the two elements of good regulation (e.g. *S* and *R*) are additively coupled. Maintaining this tight interrelation relies on independently and identically distributed (IID) statistical representations, which is optimal for generalized learning. Other representations can more accurately describe complex and even chaotic states of *S*.

How does *R* represent the system in question? Any representation must have three properties (Ashby, 1969): 1) an inventory of control responses (what to do for each encounterance), 2) decision rules that generate all possible control responses, and 3) leveraging self-organization and generating appropriate control responses. In a game theoretic formulation (Umpleby, 2008), the variety relevant to our good regulator is expressed in terms of discrete state spaces, which enables state tracking (Merrill et.al, 2024). State spaces in GRs are a mix of symbolic (tokenized) and continuous (dynamical) entities that must be consistent between *R* and *S*. One critical feature of EGRT world models is the Law of Requisite Variety, which states that there is a similar amount of variety in both *S* and *R*. This variety can be in the form of the number of tokenized states or various forms of discretization. Proper variety in *R* includes a list of responses to cover all possible behaviors of *S*. Only variety can destroy variety: the number of states (variety) in *S* must be matched by the number of states specified in *R* (Porter, 1975). Ambiguities and misclassification by *R* are the consequence of underspecifying *S*: cases in which the number of states in *S* is much greater than *R* can impede *R*'s ability to reach all possible outcomes (Casti, 1985), while cases in which the number of states available to *R* is much greater than *S* leads to oversampling.

Another aspect of good regulation is the ability to continuously monitor the response of *R* given the state of *S* over many interactions. The distance to target *G* becomes smaller as positive feedback acts as a comparator, while the lack of progress towards *G* results in negative feedback. In trying to maintain good regulation, summarized as progress towards *G*, *S* can overwhelm *R* with too much variety. Once overwhelmed, *R* is unable to act as comparator and becomes unsuited for a changing world. The ability of *S* to overwhelm *R* may also be due to insufficient initial variety. Variety in both *R* and *S* consists of an internal ensemble with discrete states. The number of discrete states allows for correspondences between discrete states in *R* and discrete states in *S*. In combination with the feedback mechanisms mentioned previously, the regulation of *R* and *S* resembles a memoryless set of discrete states. Markovian Systems in their partially observable form (Smith et.al, 2022), resemble cybernetic block models and feedback loops in that they can exhibit causal relations and non-explicit decision-making. Equivalencies between *R* and *S*, or structural similarities based on one-to-one correspondences, are structure-preserving correspondences between one the states of subdomain (*R*) and another (*S*). This is expressed in graphical form in Eq. 1.

$$f\colon\ R \rightarrow S\ ;\ R \in \square \\ f^{-1}\colon\ R \rightarrow S;\ R \in \square \qquad \textbf{[1]}$$

where $\square$ is an open set, $f$ is a feedforward relation, and $f^{-1}$ is a feedback relation. This gives us a one-to-one mapping both between states in R and S, and under feedforward ($f^{-1}$) and feedback ($f$) conditions. This provides a basis for both feedforward (forward) and feedback (inverse) relations, where *R* is an element of a closed set only. Feedback relations map to the same number of states as the feedforward relation. This gives us divergence between *R* and *S* that motivates the hypotheticals shown in Figure 2. Under ideal conditions (isomorphism), these relations are lossless and exhibit reversibility. The structure of the mapping between *S* and *R* is revealed, as well as an external goal of discovering the appropriate regulatory strategy to reach *G* (Figure 1).

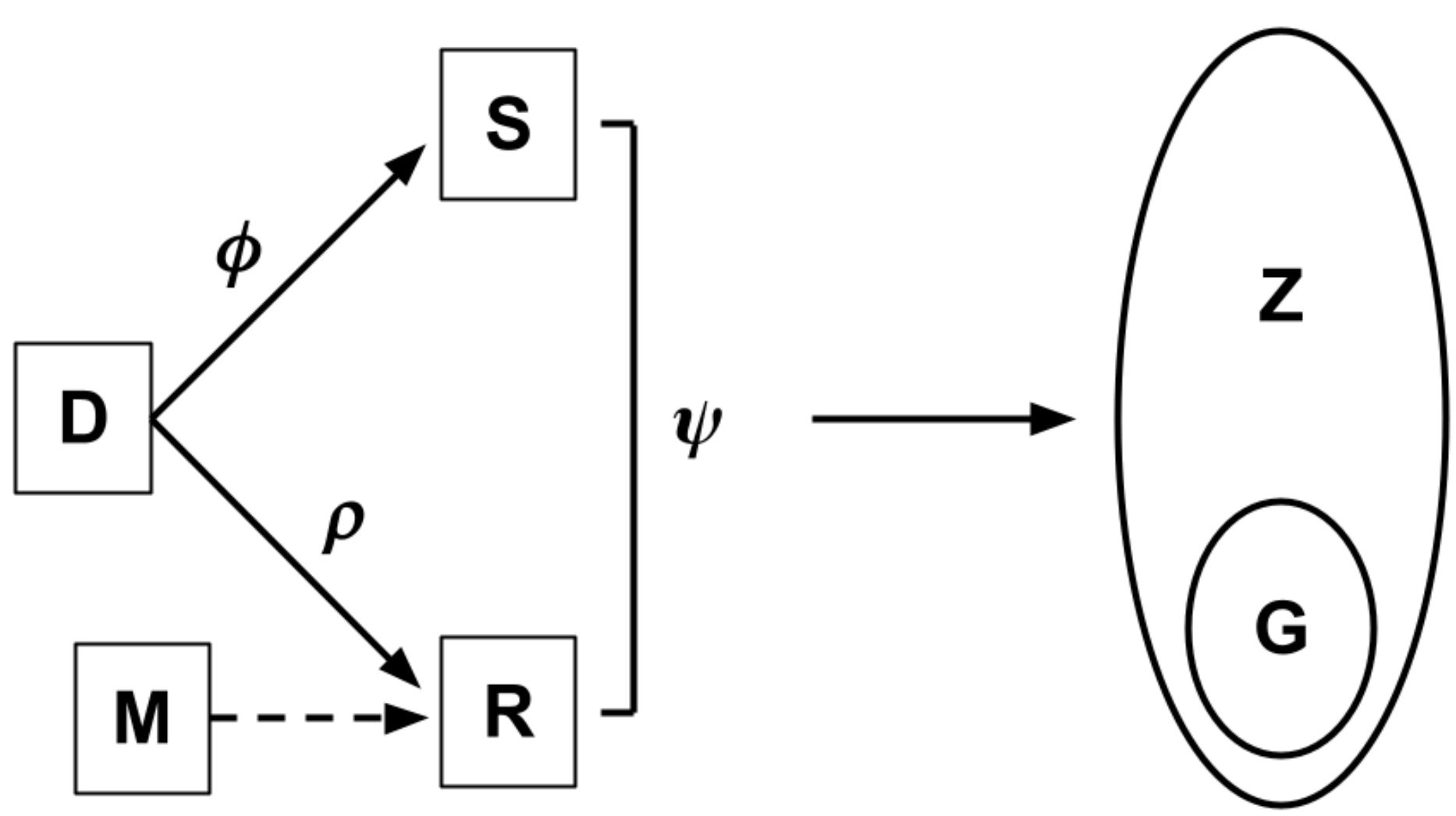

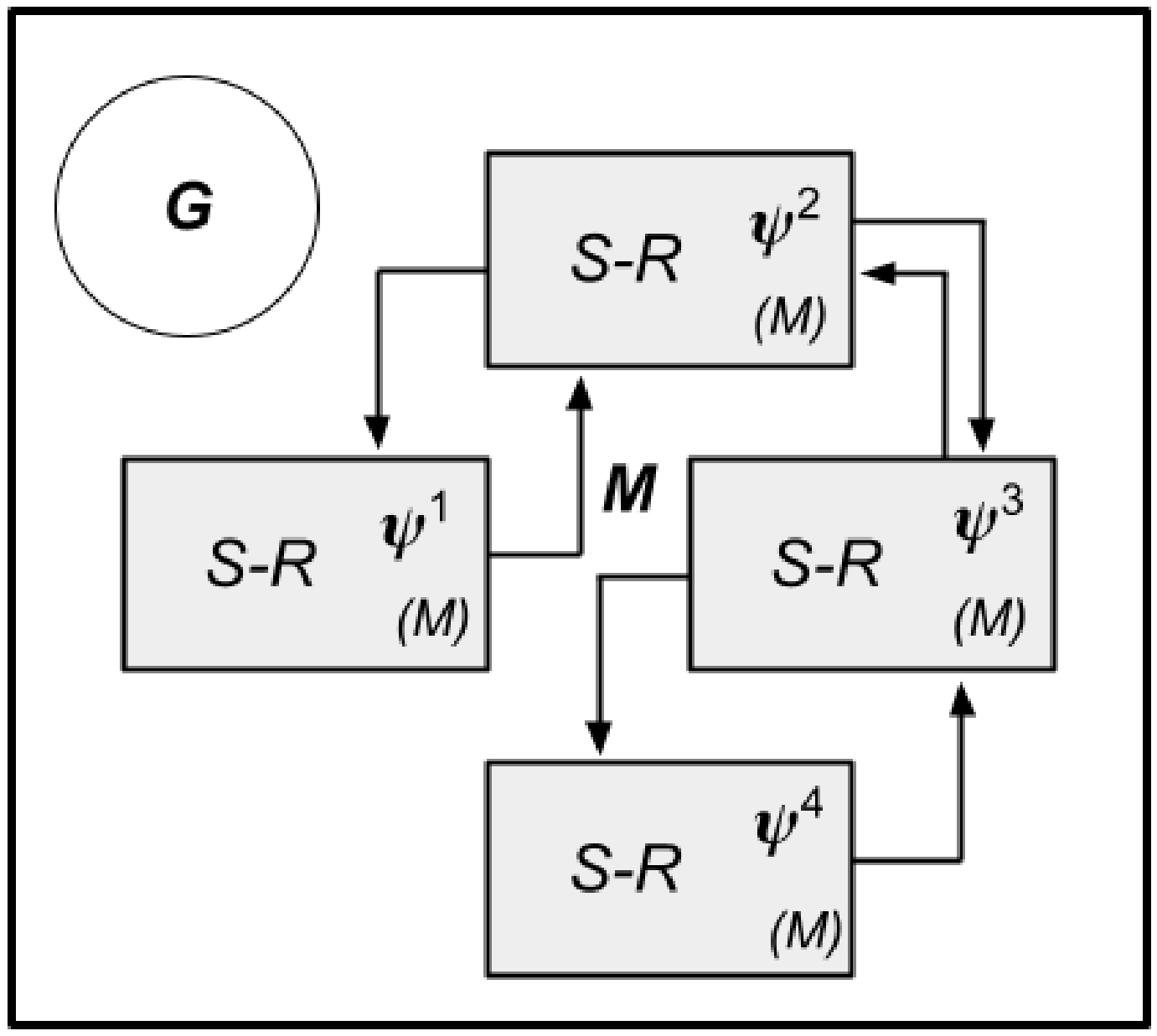

Figure 1. A: Specifying the EGRT with variables from Somerhoff (1974). The one-to-one mapping $\boldsymbol{\psi}$ describes the relationship between S and R. Disturbances $D$ to both $S$ ($\phi$) and $R$ (ρ) affect the ability of $\boldsymbol{\psi}$ to achieve the target regulation $G$ given all possible configurations $Z$. Adapted from Wikimedia user Danielsltt. B: A demonstration of closed loop regulation to reach external goal $G$ at the local and global scale. Each gray box contains a closed loop relation (closed-loop feedback supervised by internal model $M$) on local domain $\boldsymbol{\psi}_n$. All gray boxes representing components in a global closed-loop relation are supervised by global internal model $M$. Additional relationships between parameters described in Appendix A.

A regulator-system coupling is a set of reciprocal interactions (or bidirectional communication) between $R$ and $S$ on common domain $\boldsymbol{\psi}$ (Maturana, 1970). Consistent bidirectional communication between the controller and the system on this domain is a hallmark of good regulation. Importantly, the domain $\boldsymbol{\psi}$ is constantly changing as the dynamics of interactions between $S$, $R$, and $D$ unfold. Bidirectional communication allows for the acquisition of discrete states by $R$ resulting from exposure to $S$, and can be captured by Markovian dynamics. Corresponding discrete states of $R$ and $S$ collectively describe information context of the $\boldsymbol{\psi}$ domain. The mapping of states between $R$ and $S$, in addition to non-Markovian memory-dependent formulations such as feedback-dependent behavior (Narendra and Thathachar, 1974), drive learning and acquisition. With enough training, feedback provides a deep memory of state probabilities that results in behavioral norms. Rather than operating merely from weight matrices, an EGR world model provides multiple $\boldsymbol{\psi}$ domains as a global description of a system's history and future tendencies.

Figure 2 demonstrates two hypothetical scenarios where *R* tries to match and control *S*. Figure 2A features a regulator with a small number of states (3) trying to regulate a system with a larger number of states (6). This two-fold mismatch results in *R* missing some of the properties in *S* still conforms to the law of requisite variety. That is, while there is at least one state in *S* that corresponds to a state in *R*, the number of states in *S* is not so great as to destroy variety. Figure 2B demonstrates what the destruction of variety looks like: *R* contains significantly more states (20) than *S* (3). While this always ensures a corresponding state in *S*, the additional states in *R* cannot act as a comparator to *S*. This latter example not only demonstrates a poorly modeled system where *R* does a poor job of matching the content of *S*, but also provides a lesson regarding world modeling: a world model must provide access to at least as many states as a potential controller, but not so many as to introduce ambiguity. A world model connected to a machine learning (ML) algorithm provides an example. Given the diversity of a training set, a world model must represent at least as many states as the resulting latent space, but not so many as to limit feedback by overwhelming the model with endless interpretations.

### 1.2 Good Regulation as Prediction and Circularity

Good regulation also relies on a control-theoretic perspective. Control concerns how dynamical systems can be manipulated in a way that achieves desired outputs and maintains a stable system (Bechhofer, 2021). Regulation involves keeping the system dynamics constant in the face of variation. As a dynamical system, *R* approximates a desired trajectory. *S* and *R* can be treated in isolation while also being part of a feedback motif. Tracking requires two things from the regulator (*R*): measurement and response. In this way, a controller can counter disturbances and imbalances through iterative evaluation. This leads us to a formalism for closed-loop feedback: fundamental units of interactive evaluation that are equivalent with the domain $\boldsymbol{\psi}$ described in Figure 1. As a local motif that can be applied in parallel, *S*-*R* closed-loop relations is a common motif which serves as a compositional tool for building very large cybernetic systems (Capucci et.al, 2022).

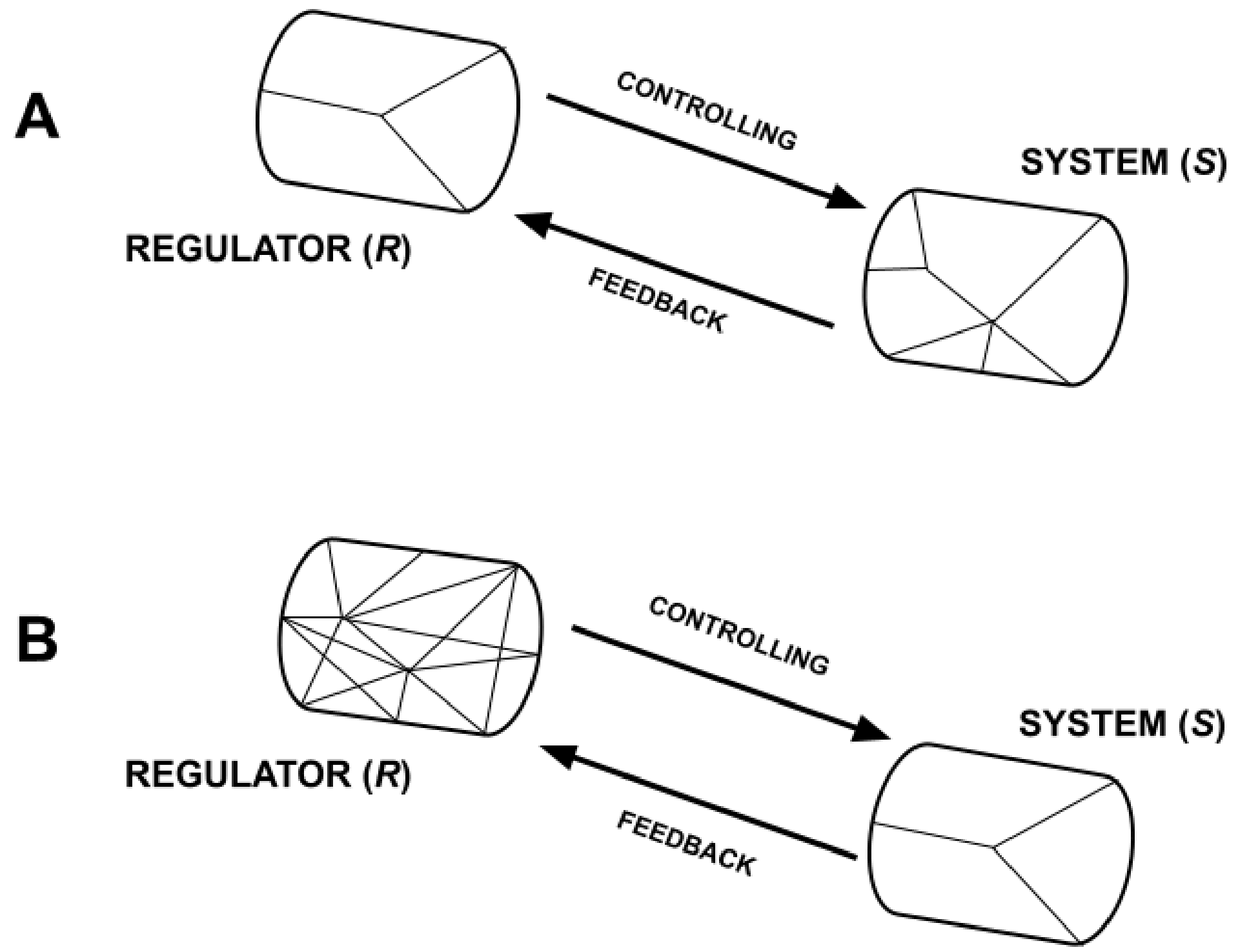

Figure 2. Capacity of regulator (*R*, denoted by A) and system (*S*, denoted by B) for two different bandwidth scenarios (near-isomorphy and aliasing). The states of *R* and *S* are represented as discrete elements, which can be generalized as embeddings that represent tokens and dynamical states. Details of figure described in Appendix B.

In cybernetic parlance, prediction results from control, or the ability to generate behaviors based on externally-imposed goals. The PID (continuous form, (Baltieri and Buckley, 2019) and related PSD (discrete form, Gu et.al, 2021) frameworks allow us to generalize our *S-R* closed-loop relation to dynamical systems. Here we will introduce PID (proportional, integrative, and derivative) regulation (Eq. 2) as relevant to thermodynamics-inspired EGR models (Bechhofer, 2021)..

$$\boldsymbol{\psi}(t) = \underbrace{\boxed{K_p\, e(t)}}_{A} + \underbrace{\boxed{\frac{1}{Ti}\int_0^t e(\boldsymbol{\tau})\, d\boldsymbol{\tau}}}_{B} + \underbrace{\boxed{T_d\, \frac{de(t)}{dt}}}_{C} \qquad \textbf{[2]}$$

PID defines the idealized one-to-one relationship between *S* and *R* as the error value between target values and the observed state of a given system. In Eq. 2 $\boldsymbol{\psi}(t)$ describes the domain of an S-R closed-loop relation over time, and Eq. 2 describes the proportional (Eq. 2A), integrative (Eq. 2B), and derivative (Eq. 2C) components, respectively. Applied to the EGR, PID describes the duration *R* is able to withstand being mismatched to a given setpoint. The proportional aspect measures the present error value, while the integrative aspect aggregates past error values. *P* and *I* construct a model for *D*, which does not generalize outside of constraints *P* and *I*. Conflicting objectives can introduce additional difficulties. Nevertheless, PID models open up the possibility of optimizing systems for one-to-one correspondence, thus meeting the most restrictive criterion of good regulation. Baltieri and Buckley (2019) connect the PID model to cognitive functions such as motor control, learning, and attention.

Circular causality describes the temporal unfolding of relations between regulated and regulating variables in a feedback loop. The circular causality resulting from a *S-R* closed-loop relation provides a means to understand the dynamics of real-world systems and world models. Bateson (1972) provides an account that involves evaluating multiple equivalent plausible relationships arising from interactions between *R* and *S* (Figure 2). *R* can regulate *S* in several different (and perhaps conflicting) ways. Patten and Odum (1981) propose that systems with feedback are a special type of cause and effect, where individual inputs occasionally produce large effects from small feedbacks over time. Accumulation of feedback over time produced a response disproportionate to the current signal. In terms of Figure 1, causality is viewed as a disturbance (*D*) to *R* (ρ), serving to temporarily mismatch *R* and *S*, and testing the ability of a given *S-R* closed-loop relation to match each other's behavior. While temporary decoupling of R and S can occur, such responses are expected to be unpredictable and transient. Constructing a world model that responds to local conditions provides a means for *R* to match *S* via individual *S-R* closed-loop relations, even under high levels of disturbance. Causal relationships between *R* and *S* require *R* to produce behaviors that predict target *G* on domain *Z* (Figure 1). The behavior of *R* is a restricted solution on domain *Z*. The resulting teleological behavior can often be spurious. Pareidolia in deep learning models (Gupta and Dobs, 2025) are an example of this: *R* locks on to a target solution in *Z* that looks like a pattern but does not match *G*. *S* can constrain *R* in ways that lead the system to an externally-specified goal, namely the target *G*.

### 1.3 Properly Understanding the Behaviors of *R*

How do we assure that our world model does not live in an alternate universe of misattribute in a manner that encourages model hallucination? Before applying the closed-loop feedback construct to actual systems, we must clarify the multitude implications of good regulation. Rosenbleuth and Weiner (1950) define the goal-directedness of a cybernetic agent as adaptive behavior controlled by negative feedback, which is consistent with modern notions of RL and cognitive control. While functional analyses might imply goal-directed teleology (Simon, 1976), lawful relationships do the real work of shaping *R*'s behavior (Canfield, 1964). Teleological assumptions may also incorrectly imply spurious causation: prediction (and ultimately control) should be neutral with respect to a goal. Gruner (1966) argues that function need not imply teleology. In fact, we can distinguish between cases in which a goal is consciously pursued (internal goal-directedness) and cases in which a goal is not consciously pursued (internal or external goal-tendencies). In cases where we utilize *M* as a second-order cybernetic referee, we might expect that active teleology plays a greater role. Indeed this is where we can benefit (or harmed) by biased and bounded behavior, which affect the degree to which data is locally-sampled and sensorimotor affordances are relied upon. Yet we propose that even when second-order observation is distributed locally (Figure 1B), cybernetic regulation can proceed at the global scale that *does not* operate according to the same assumptions as second-order cybernetic regulatory function. This is due to interactions between local second-order closed-loop feedback motifs, leading to applications in an open-loop world model setting.

One viable alternative to cybernetic teleology involves self-reflective control using a version of the Ashby box (Ashby, 1954): an elementary non-trivial machine that demonstrates the *S*-*R* closed loop motif. Autonomously, an Ashby box constructs, deconstructs, and reconstructs self-function (Cretu, 2020; Ashby, 2020; Ashby, 2022) in a manner similar to diffusion models. The Ashby Box itself is a second-order cybernetic model (Maturana, 1970, Scott, 2004; Jeon, 2020; Cretu, 2020), which means that *R* is an observer of its own behavior (Scott, 2004). Not only does *R* have access to *S*, but also *R*'s performance on domain *Z*. Ashby (1972) defines the setting of internal goals as the observation of adaptive behavior. Ashby boxes provide a self-referential signal to enable a form of self-supervised learning. This stands in contrast to ascribing motivations or agentive imperatives to the regulator, but also suggests that a properly configured world model provides a means towards improved learning performance. Rather than *R* acting intentionally, *R* matches the information content of *S* by reciprocating *S*'s behavior, assuming that both *S* and *R* possess complement states.

### 1.4 Model Versus Instantiation

When applying the Every Good Regulator Theorem to modern AI models, we must distinguish what holds true for the cybernetic model as a generalized prediction and how they perform in a real-world instantiation. The classic cybernetics literature is heavily theoretical, which results in a poor distinction between different types of application domains (e.g. biological, cognitive, or engineered). In general, we will be referring to the generalized regulatory behavior of a cybernetic model. This can be applied broadly, and on its own does not make specific predictions. Later, we will make more explicit connections to modern AI models. The parallels we draw will not be based on formal experiments or benchmarks. This reduces our ability to make strong statements about the suitability of EGR-inspired systems for real-world applications.

We also discuss the potential of physics-based world models based on lawful behavior of statistical mechanical summaries. Modern AI models are driven by optimization (e.g. error minimization) of a latent space or landscape. Effective world models based on the EGR must possess an internal representation that is consistent with the optimization objectives for the modern AI system in question. Fortunately, statistical mechanical descriptions focus on several regularities that characterize a wide range of systems. From the standpoint of allostatic control, ergodicity is one way we can connect cybernetic regulation with optimization objectives. Given enough information about the real-world system, ergodicity involves making an estimation of a system's behavior from a sampling of the local set of inputs. The statistical mechanical approach, consistent with approaches to cognition such as the free energy principle, relies on energy minimization landscapes and their connection to the optimization criteria of external goals. While we do not offer concrete examples of how the potential advantages of EGR world models can be implemented, we do open the door for future applications.

Of course, much of the data a modern AI system will encounter is aperiodic, non-normal, or simply structureless. Multiple instantiations of world models (Del Ser et.al, 2025; Hafner et.al, 2025) have involved using semantic components or conceptual schema to contextualize data from the external world. Our proposed EGR world model takes a different approach, focusing on summarizing the statistical properties of the real-world. This allows us to augment the generalized cybernetic description and potentially link it to data-driven models (e.g. modern AI models). The variable nature of real-world data is why we focus on the temporal criticality and alternating conditions examples: providing a means to adapt to such irregularities is a key aspect of model robustness and adaptability in a wide range of application domains.

To better understand the goals and teleological trends that typify cybernetic systems, consider the emerging field of machine behavior. Proximate machine behaviors are observationally triggered and generated in specific settings (Rahwan et.al, 2019). Machine behavioral mechanisms come in the form of simple rules or algorithms. External inputs contribute to an internal model of the world, setting up both externally-driven optimization and internally-driven regulation. The former is typical of modern machine learning, while the latter can be described by allostasis. A world model based on the EGR utilizes a combination of *S*-*R* (first-order) and *S*-*R*-*M* (second-order) relations to improve the performance of a given modern AI system. The output of these behavioral mechanisms over time, particularly in the context of matching internal goal states, produces adaptive and robust outputs. Let us consider a fairly simple benchmark task: the pole-balancing problem. Pole balancing can be achieved with a first-order control system or Reinforcement Learning (RL) algorithm. Yet EGR world models augment first-order control with the ability to overcome perturbations, noise, and entirely new contexts.

## 2. Dynamical and Learning Paradigms Challenge Idealized Good Regulation

In this section, we will provide three examples of how the EGR can approximate regulation of complex learned behavior: temporal criticality using an avalanche model of criticality, non-normal denoising, and procedural learning in alternating environments. This provides context into how an EGRT-style regulatory system can be applied to world model prediction. As all three examples are based on physics, they provide a nice example of EGRT compliant regulation. Thermodynamics-inspired EGR models are particularly well-suited for world models. Statistical mechanical regularities allow for compact representations, which are sufficient for a complete and well-defined problem description (Chollet, 2021). Learning from statistical regularities enables broadly generalizable skill learning. Learning skills that generalize over multiple classes of tasks provide so-called deep statistics, which enhances robustness and stands in contrast with the surface statistics (e.g. textures) typical of deep learning models (Chollet, 2021).

### 2.1 Sandpile avalanches and diffusive processes challenge good regulation

One way to test the notion of good regulation is to challenge it with a stochastic, non-uniform process. The Abelian sandpile model (Dhar, 2006; Fey et.al, 2010) can be used to illustrate how a good regulator can produce and respond to a perturbation. Abelian sandpiles[7] are defined by burstiness and unpredictability, accumulating over time until they become unstable. The resulting instability is resolved as a series of avalanches, the signature of which is scale-free (distributed over many different timescales and orders of magnitude). We propose an avalanche model of good regulation where *S* provides a feedforward signal to *R* that accumulates and transmits this signal at random time intervals. This introduces bursts of information to *R* that resemble an avalanche. Each avalanche drives *R* away from equilibrium. This behavior produces avalanches of different sizes over time. When enough time elapses between avalanche events, *R* can rediscover its setpoint and restore equilibrium by adding or subtracting corrections (additional details in Appendix C).

The avalanche model of good regulation is relevant to time-dependent ML models. The distribution of avalanches over time resembles an autoregressive fractionally integrated moving average (ARFIMA) model (Granger and Joyeux, 1980; Taqqu et.al, 1995). Once again, statistical structure is critical to the development of physics-based world models for modern AI systems along with the EGR more generally. For example, the DeepSeek (diffusion-based LLM) consortium makes a claim that an LLM post-trained with an RL-based reasoning framework exhibits behaviors analogous to *aha!* moments in human cognition (DeepSeek consortium, 2025). *Aha!* moments are instances of recognition based on a long sequence of prior information. While this claim might be controversial, another group (Stetchly et.al, 2025) observes such behaviors under two conditions. Sequential learning based on the assembly of tokens. can lead to this phenomenon. Identifying sudden moments of composition also occurs in the absence of an internal model for recognition (Stetchly et.al, 2025). This suggests sudden moments might be critical in nature, relying on the structure of the data (non-normally distributed) rather than gradually piecing together reasoning step-by-step. Indeed, natural language exhibits long-tail statistical signatures. Even for randomly generated text, related words and concepts are acquired as a group while critical components are acquired at much lower frequencies (Li, 1992). Diffusion-based language models increase performance on a wide range of tasks (Deschenaux,

[7] Sandpiles are a toy model that serves as the source of avalanches at non-deterministic time intervals. Abelian dynamics refers to the unordered nature of grains in a given sandpile: the relative order of each addition to the sandpile does not affect the frequency or size of a resulting avalanche.

2024; Nie et.al, 2025) while sharing features with our avalanche model that could be complemented by a physics-based world model. Mapping this from LLMs to EGRs involves determining whether selectively delayed feedback between *R* and *S* explains observed behavior better than continuous feedback between *S* and *M* (first-order cybernetics) and delayed communication between *M* and *R* (second-order cybernetics).

Figure 3A-3C demonstrates avalanches generated from a *S-R* open-loop relation. Figure 3D-3M demonstrates a forward diffusion model where masking with noise resembles generalized learning. Figures 3A-3C shows acquisition to R from S using simulated data for $10^3$ moments where each moment is a random value (0 → 1). The feedforward component is automatic for the random diffusion case, and accumulation at random intervals (4 → 10) in the non-random diffusion case. Using a simulated example, avalanches serve as ρ moments (Figure 3A, blue) allow for a comparator to learn a distribution. The rank-order plots (Figure 3B-3C) set the ability to learn a strategy for recovery to an equilibrium point. For the forward diffusion model, the α parameter is varied for the random (D-H: 0.2 → 0.9) and non-random (I-M: 0.8 → 0.01) case. This experiment shows a sequential process translated into characterization by statistical ensemble, providing a description for a potential world model. The demonstration of diffusion shown in Figure 3D-3M is consistent with modern diffusion models in ML. Given a training set of images, noise is progressively added until a full white noise mask for each image is achieved. The original image is then recovered, with the noise providing variety for the diffusion model. This can be stated in terms of the EGRT: *S* sends a progressively noisy signal to *R*, which is corrected through feedback from *S*.

Azizan et.al (2024) explore the outcomes of stochastic gradient descent for generalized non-convex problems such as out-of-distribution data. This provides atypical instances that lie outside of the normal statistical distribution. In EGRT terms, this provides a challenge to *R* based on a closed set which challenges *R* to incorporate newly encountered properties of *S*. While using an open set approach would be preferable (Yang et.al, 2023), it is not always sufficient. While using an open set approach would be preferable (Yang et.al, 2023), it is not always sufficient. In the case of world modeling, the discovery of novel configurations of *R* and *S* (not simply representing *R* as an open set) is key. One way to overcome this limitation is to use a Boltzmann-Gibbs[8] distributed gradient descent process that visits critical regions more often than other regions of problem space (Azizan et.al, 2024). We propose a world model of forward denoising that includes both Gaussian and Power law distributions of noise (Figures 3I-3M) that might facilitate gradient descent in a way that speeds up learning. In the context of an *S-R* closed loop relation, unpredictable shifts in control might be analogous to making an emergency stop while cycling: the detection of this occurrence not only disrupts good regulation, but in fact *cannot* be deterministically prepared for. As with the previous examples from LLMs, our forward noising examples (Figures 3I-3M) are analogous to Aha! experiences, where a sudden jump in recognition leads to instantaneous problem-solving success (Wiley and Danek, 2024).

[8] A Boltzmann-Gibbs distribution is derived from statistical mechanics, and describes a probability distribution that provides an estimate of when a given system will be in any one state as defined by its energetic profile.

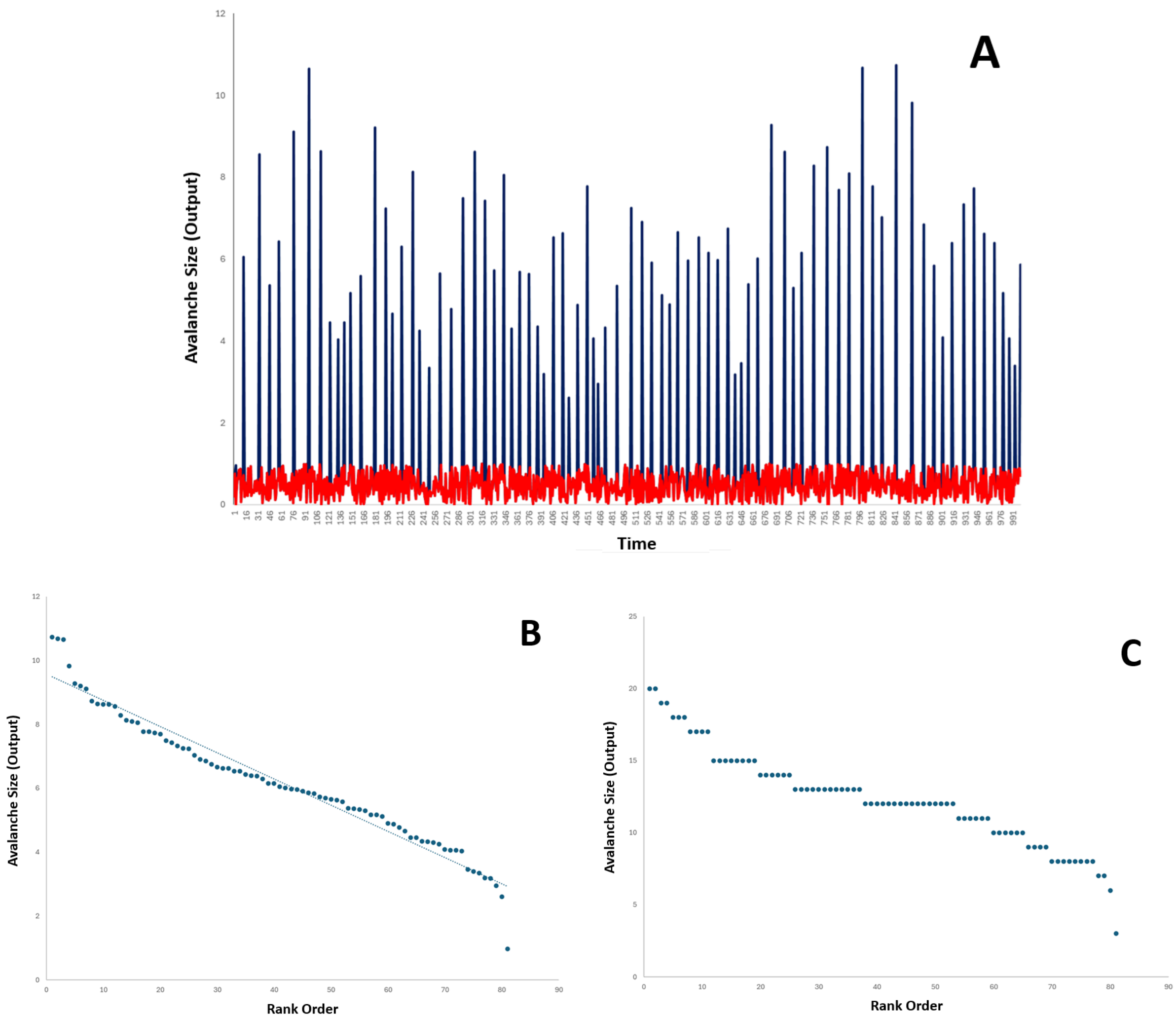

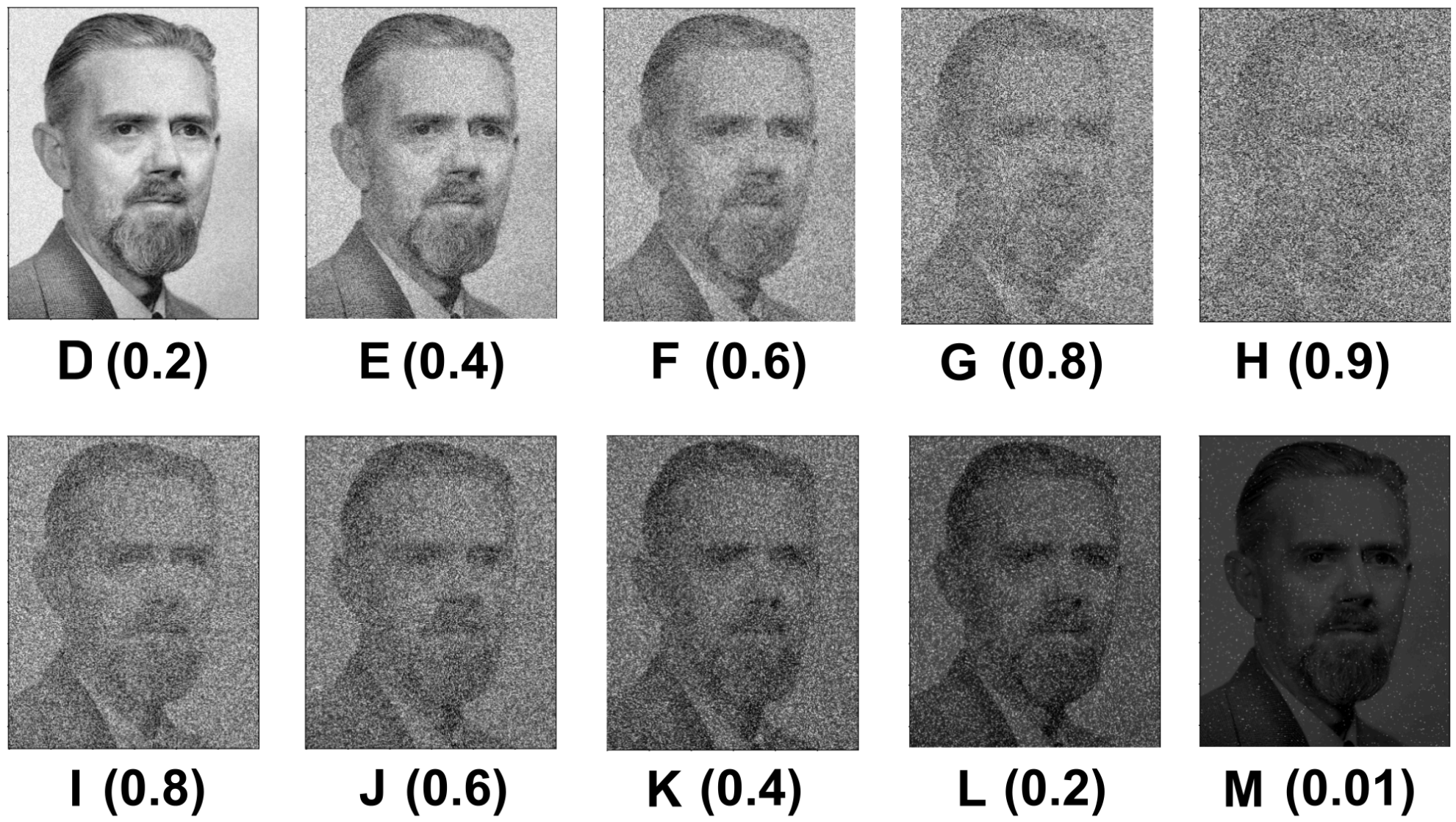

Figure 3. A: time-series plot of avalanche size for random (red) and non-random (blue) diffusion; B: distribution of avalanche magnitudes, rank order from largest to smallest; C: distribution of intervals between avalanches, rank order from largest to smallest. $t$ = 1001 moments for 81 avalanches. D-H: image of W. Ross Ashby progressively embedded in random noise ($\alpha$ value in parentheses). I-M: image of W. Ross Ashby embedded in progressively sharper power noise (shape parameter ɑ in parentheses). Methodological details for these simulations are described in Appendix C.

### 2.2 Alternating procedural acquisition challenges good regulation

Another way to test the notion of good regulation is to challenge procedural learning with alternating environmental conditions. Procedural learning involves acquiring skills that become automatic once encoded in memory (Hong et.al, 2019). Procedural learning is a challenge to embodied robotics research, where a world model must approximate continuous fluctuations and circular causality in the physical environment (Vernon et.al, 2015). Acquisition is facilitated through the introduction of alternating conditions, such as directional force fields, inertial forces, and sequences of presentation (Shadmehr and Mussa-Ivaldi, 1994). These alternating conditions serve to generalize conditions for the learner, and provide a basis for discriminative learning (Figure 4). The learning-unlearning-relearning paradigm[9] shows how this works: consider an embodied agent that rotates 90° for every learning phase. That agent experiences the system both in-phase (0° and 180°) and anti-phase (90° and 270°). When positioned towards 0°, the mapping $\boldsymbol{\psi}$ is exposed to a rotational force field. When rotated 90°, anti-phase information distorts the force field and conflicts with in-phase information. This frame of reference alteration interferes with the previous learning epoch. Added variety of a conflicting type partially destroys the original mapping ($\boldsymbol{\psi}$) between *S* and *R*. An additional shift to 180° returns the agent to an in-phase orientation, where traces of the original in-phase mapping is reinforced ($\boldsymbol{\psi}$'). Adding additional rotations enables global sampling of variation in the system necessary for turbulent conditions over time (Mussa-Ivaldi and Bizzi, 2000). A NeurIPS MyoChallenge winner (Chiappa et.al, 2024) used two phases (learning-relearning) to achieve alternating procedural learning. Counterclockwise rotations were followed by a mixing clockwise and lack of rotations. Increasing variation in the second phase forces the learning model to engage in inference (Chiappa et.al, 2024). In a more general sense, the learning-unlearning-relearning approach forces a temporary mismatch between *R* and *S* that serves as a mechanism for learning. As in the case of criticality, *R* is faced with a challenge by new scenarios provided by *S*, which can be overcome with sufficient variety.

[9] Having originated in the motor learning literature, unlearning is also a key aspect of memory consolidation in Hopfield Networks, which are recurrent neural network models of associative learning. The Hopfield Network unlearning process is conceptualized as a dreaming period during the learning process, which serves to optimize performance.

This procedural learning example is exemplified in Figure 4, which shows a sequence of hypothetical rotational force fields (Figure 4A) and CMYK color space (Figure 4B). In the case of rotational force fields, *R* must match the difference between each force field to counter the variety of *S*, general enough to realize closed-loop motor control. This agent explores a triangular color space by allowing a closed-loop feedback model of the system to act as a comparator that can sharpen distinctions between points on a color, creating a gradient of colors as a sequence of states. The comparator takes the feedback instance (newly encountered CYMK value) and contrasts it with previously encountered values stored in *R*. The feedforward signal is then a contrast (a perturbation ρ to *R*'s model) and biases *R* towards the next location in *S* to explore. In Figure 4 (bottom), the gradient is learned from an initial mapping between *S* and *R* over subset *G* of CYMK values, then unlearned as a larger subsystem (*G'*, more CYMK values) in the color space is sampled. Relearning is where the subsystem expands even further: ideally, relearning occurs as a resampling of the previously encountered CMYK space. The subsystem then expands to encompass the entire CYMK space (*G''*).

Alternating procedural learning produces dynamical movement behavior characterized as sensorimotor learning under shifting environmental conditions (Bennequin et.al, 2009). Therefore, *S*-*R* closed loop relations are local for a three state sequence, but global for longer sequences. Alternating procedural learning for these longer sequences provides a means to learn color spectra, in particular the gradations between color values in a CMYK color space (Figure 4B). Procedural learning (Figure 4) is not only applicable to force adaptation and color acquisition. So-called contrastive learning techniques operate in the same manner, providing negative examples to balance out positive examples of a given class, thus reinforcing the boundaries of a category or class. As informative contrasts are limited, Robinson et.al (2021) provides a technique for contrastive learning using hard negative examples where positive samples are intentionally pushed apart by hard negative signals. The resulting clusters are learned from the introduction of *1/f* or superdiffusive noise. Similarly, the regularization method called stochastic noise injection introduces asymmetric exponential noise (Camuto et.al, 2021).

A learning-unlearning-relearning paradigm might improve informativeness of the world model by taking advantage of representing heterogeneous and expansive non-IID network structures.

## 3. World Models and Intelligent Behavior

### 3.1 World Models Are Consistent With Good Regulation

One way to approach physics-inspired world modeling according to the EGR involves achieving continually improved matching ability between *R* and *S*. A world model incorporating the EGRT can bridge this gap. While the *S*-*R* closed loop relation is generic, *R* can match *S* under diverse environmental conditions. We can also think about unaligned or poorly-performing aspects of contemporary generative systems as a misalignment in terms of runaway negative (hallucination, selective breakdown, false memory, lack of incorporation) and positive (creativity, self-referential loops, transcending self-referential loops, and underspecified input data) regulatory components (Chollet, 2019). Application of regression techniques to unsuitable problem domains demonstrate the dangers of runaway positive feedback: high-dimensional polynomial regression leads to extreme overfitting, and thus no generalization. Yet good regulation is more than simply interpretability or improved efficiency: a compact generative model that is robust to heterogeneous structure and universal mechanisms (all of which describe the *S*-*R* closed loop relation) is also required

### 3.2 Connecting the EGRT to Contemporary AI Models

Now we want to apply the various aspects of good regulation and the EGRT to contemporary AI models. The difference between *R* and *S* is the difference between the model of the world and the training dataset. To arrive at conclusions about world modeling, we must consider different styles of modeling, their state-of-the-art in the ML/RL literature, and their connection to various representations. In training physics-inspired models, predicting stochastic and potentially chaotic systems is paramount. GRs provide a means to escape this conundrum by enabling complex regulation. If a given model is able to capture these divergent states (maintain ergodicity[10]), then prediction becomes possible. Non-equilibrium distributions require open systems that resemble Markov processes reconfigured according to the rate of change in entropy (Pollard, 2016). Similarly, out-of-distribution sampling in dynamical systems allows us to generalize across attractor basins, being able to find the proper context in unseen domains (Goring et.al, 2024). This approach to out-of-generalization allows us to better forecast extreme events, anticipate tipping points, and predict post-tipping point events (Camuto et.al, 2021).

The EGRT shares structure with ML model training, latent space features, and other potential methods for constructing AI-compatible world models. Generative Flow (GFlow) networks (Bengio et.al, 2022)[11] sample from a diverse training set during active learning, providing a flexible and open representational model. Aside from model performance on specific tasks, GFlow nets estimate statistical distributions for heterogeneous and unspecified components of a potential world model. As with Markov processes, the structure of GFlow networks allow discrete supergraphs to be constructed from subgraphs. This reveals a shortcoming of simply scaling up the *S-R* closed loop relation. As an alternative to GFlow networks, Generative Simultaneous Localization And Mapping (G-SLAM – Safron et.al, 2022) allows for the expansion of neural representations via latent space sampling to perform sensorimotor integration. G-SLAM architectures[12] expand by discovering potential loop-closures, graph relaxation, and node duplications for enriching representation. An EGR world model features single *S-R* closed loop relations that serve as a low-dimensional source of regulation, while multiple *S-R* relations deployed in parallel that manage feedback and evaluation that responds to heterogeneity inherent in the real world. While model stitching can enable the convergence of representational capacity between local and global scales (Bansal et.al, 2021), the EGRT may enable more sophisticated techniques.

[10] Ergodicity is achieved when the average behavior of a system is equivalent to all other examples of the system (average behavior predominates). That system will eventually explore all possible states given a sufficient amount of time.

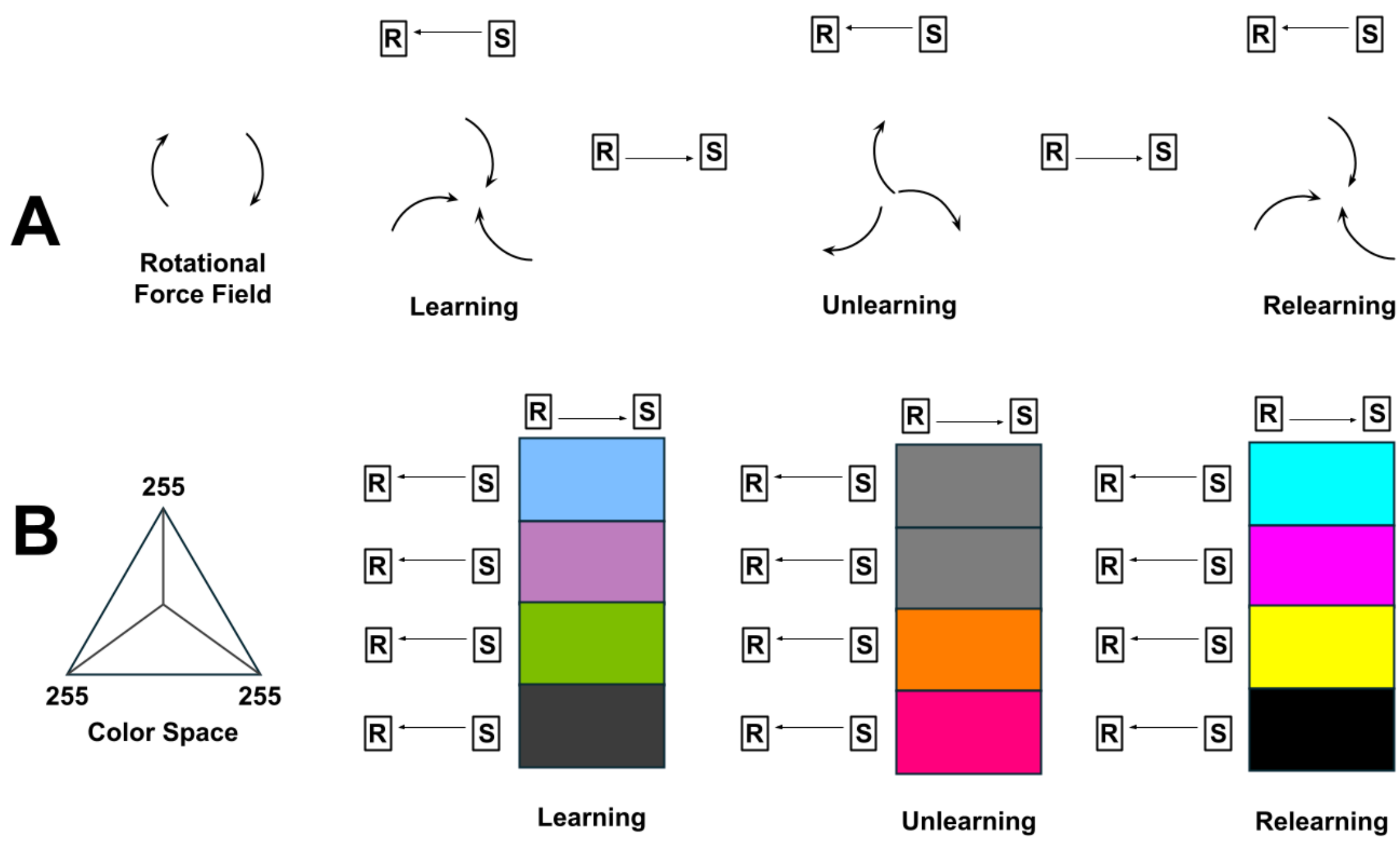

Figure 4. The learning-unlearning-relearning paradigm for rotational force fields (top) and the CMYK color spectra (bottom). Components of the *S-R* closed-loop relation activated at a certain

[11] GFlow networks combine reinforcement learning and variational learning to implement system 2 inductive biases in a deep learning framework. Variational learning allows for maximum likelihood estimations in cases where the underlying distribution is complex and partially unknown. In combination with a reward signal, this enables slow, deliberate, and intentional behaviors associated with system 2 cognition.

[12] G-SLAM is utilized for obstacle avoidance and navigation. Simultaneous localization is defined by a map of points in a continuous space. Each point is encountered by an agent at a given probability, and these probabilities can be adjusted by the adaptive (generative) behavior of the agent over time.

point in the sequence are shown as an arrow from $S \rightarrow R$ or from $R \rightarrow S$. Additional details are described in Appendix D.

Viewing machine intelligence in terms of the EGRT allows for active representations of the input data. Intelligence is not a camera with a finite capacity, but a regulatory domain that connects different domains of knowledge. Limitations of capacity exist in the latter case, but do not limit our perspective of the world. An EGRT-based world model can utilize an observer (consistent with second-order cybernetics: Section 1) rather than supervision via labels. Applying the EGRT model to deep learning is consistent with the observations of Bengio et.al, (2009): each level of supervised deep learning architectures provide opportunities for composing a world model out of component parts. This suggests *S-R* closed-loop relations can likewise be composed into multiple *S* and *R* components (e.g. $S_1$-$R_1$-$R_2$-$S_2$-$R_3$ semi-open loop relation). Despite exploiting structural composition, truly intelligent systems require an embodied model of the system observer (Safron, 2021) to achieve sufficiently good regulation. Indeed, RL-based world model performance suggests that multiple aspects of cognition must be incorporated into machine performance (Hafner et.al, 2025). The EGRT thus serves as an abstract analogy to intelligence: is embodied information processing defined by good regulation, or something different?

## 4. Broader Implications of Good Regulation

### 4.1 What are the structural insights of good regulation?

To better understand the structure of good regulation, we can analyze the examples from Conant (1969). In Figure 1, *R* and *S* serve to jointly regulate the output *Z* so that the target *G* is achieved. *R* regulates against *S* in a manner similar to an generative adversarial system[13] (Goodfellow et.al, 2016). *R* and *S* are joint contributions of output *Z*, but in cases where *R* and *S* form a bidirectional interaction (e.g. feedback loop), *R* must provide a correction to *S*. The regulator is subject to thermodynamic and informational entropy at differing rates, which requires a continual updating of the model with regard to the environment. Taking this further, Stone and Alicea (2015) provide a model called the Cybernetic Convolutional Architecture (CCA), which places the EGRT model-system mapping into a model of suboptimality. CCA models compose single *S-R* closed loop relations to produce behaviors at different scales in a manner similar to GFlow networks and G-SLAM. More generally, a compositional approach where the global system is decomposed in a set of connected subsystems (multiple *S* controlled by multiple *R*), closely approximates the notion of a world model. Feedback is crucial to good regulation, as feedforward control disallows the accumulation of regulatory signals (see

[13] Generative adversarial systems allow for a neural network to be trained using a main training set that is contrasted with an adversarial training set. Adversarial examples might include sources of noise, augmented samples, or otherwise distorted samples. The adversarial approach can improve learning performance through positive feedback, where the adversarial samples reinforce output behaviors the model should avoid.

Appendix B). Translated to training sets in learning models, temporal or sequential structure is required as input for good regulation of the output.

Conant (1969) points to two types of discrete regulation: point regulation, good regulation results from the imposition of constancy for a scalar, and path regulation, which acts to minimize the unpredictability of sequential outcomes. The thermostat's set point is an example of point regulation. In terms of Figure 1, *R* acts to minimize the entropy (fluctuation) of *Z*, and increase the chances of reaching target *G*. In terms of path regulation, the entropy of a sequence (e.g. time-series) resulting from output *Z* is minimized, thus increasing the chances of reaching target sequence *G*. We might say that path regulation is a necessary condition for point regulation. From a second-order cybernetics perspective (Maturana, 1970), this network is influenced by changes in the system, whether they be single events or phases of behavior. This provides a route to procedural learning and subsequent predictive capacity without relying on a contrived teleological mechanism.

### 4.2 What are the limits of the EGRT?

To understand the challenges posed by complex systems (often non-IID) contexts, we can use two types of data that may pose challenges to statistical and procedural learning, respectively: avalanche frequency for criticality and alternating between different forces and learning conditions. *R* will develop a generalized strategy to deal with constant and irregular change patterned in different ways. For example, an agent is trained using a combination of data collected by other agents to learn generalizable skills like robotic walking or grasping. We can also map gradient descent (ML) and policy refinement (RL) to the EGRT variables described in Figure 3 and Appendix A in Figure 5. While we do not approximate LLM architectures in Figure 5, control theory has been used to analyze LLM architectures in a manner similar to what we are proposing for understanding and deploying good regulation (Bhargava et.al, 2023).

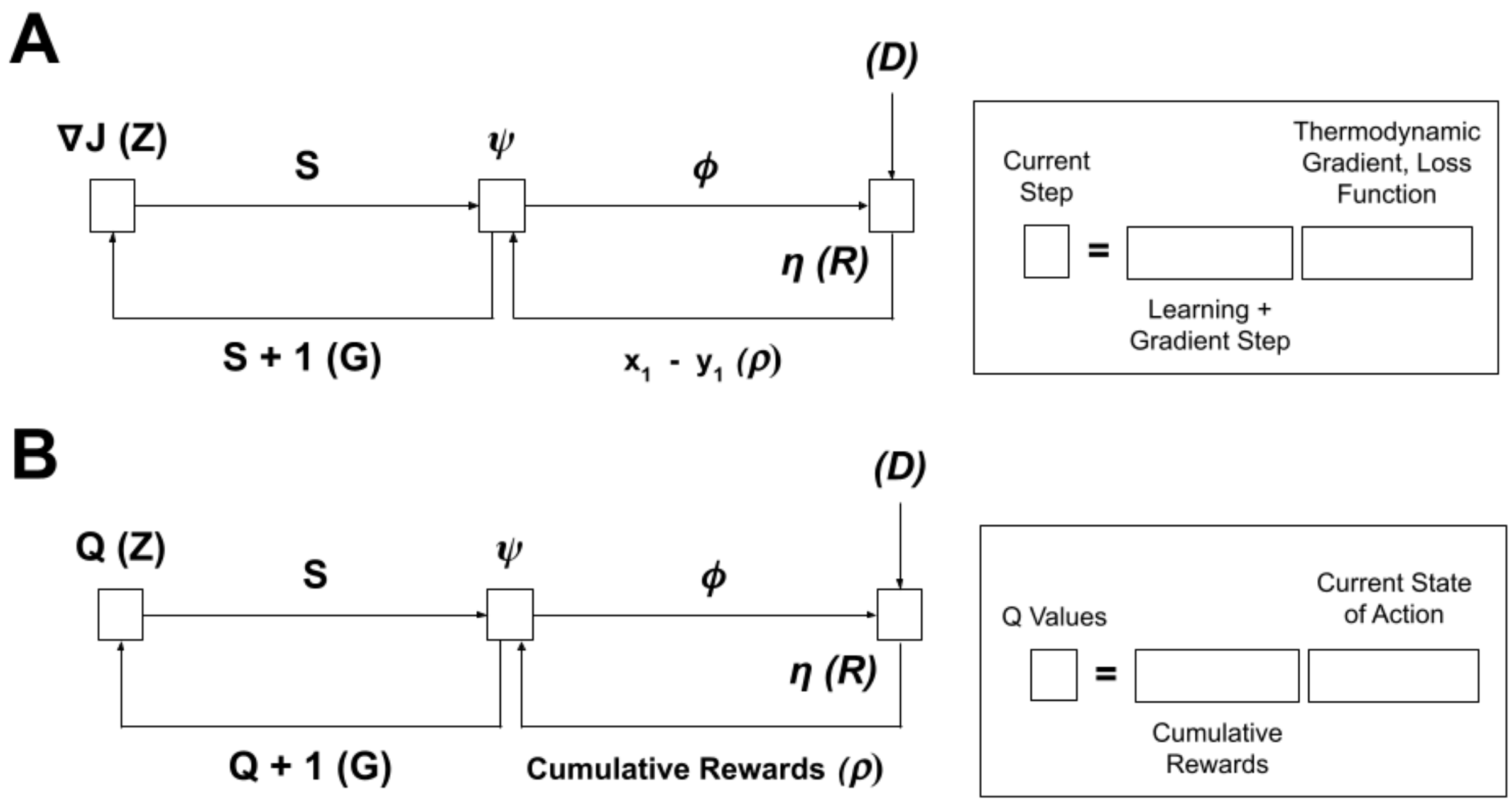

Figure 5. Examples of gradient descent (A) and policy refinement using Q-learning[14] (B) as GRs. Breaking down equation components into feedback models for gradient descent (A, left) and Q-learning (B, left). Functional model components for gradient descent (A, right) and Q-learning (B, right).

Another route to GRs that incorporate changing conditions is to achieve ergodicity based on full experience with the system. Most systems involve shifting environmental conditions that characterize novel dynamical states. Models of those systems are thus non-ergodic (Sennesh et.al, 2022) in that they require mean behaviors that approximate global states of the system. A Boltzmann statistical mechanical interpretation exists where R partitions S into discrete states, while *M* approximates time-averaged probabilities and weak ergodicity is assumed for only a fraction of states in *S* (Werndl and Frigg, 2015; Frigg et.al, 2020). Temporal criticality is similar in this respect: *R* must have experience with avalanche-scale events despite the most common state being a non-avalanche. While oversampling non-avalanche states provides us with a sufficient model of equilibrium states (Wendl and Frigg, 2015), the out-of-equilibrium nature of avalanche states leads to the failure of good regulation. Thus, map *R* becomes better than territory *S* (Hoel, 2017). Recognizing avalanches as a distribution of states comes at the cost of losing the proper temporal placement of avalanches.

## 5 Conclusion

[14] Policy refinement and Q-learning are techniques from the Reinforcement Learning literature.The Q-value is the *quality* of an action taken in a given state. The Q-learning process rewards optimal *action-selection* policies (states and actions selected on the basis of maximal payoff) by rewarding select Markov decision processes.

How suitable is the EGRT to serve as a component of modern AI systems, particularly in terms of providing a world model? We can turn to one definition of intelligence that encapsulates this potential. Wang (2019) defines intelligence as a lifelong learning and adaptation process, which occurs in real-time under often insufficient informational conditions, and is embedded in an ever-changing, non-repeatable world. In this paper, we recast the EGRT as a regulatory mechanism that captures many of these properties in the context of physical paradigms that exhibit statistical regularities. Moreover, the relations between *R*, *S*, and *M* provide a means to construct world models that not only faithfully represent the task, but offer a means to supervise (observe) the construction and update a suitable representation. The EGRT also allows us to construct an open-loop set of relations that incorporate new concepts and features. For example, the observations of $R \in \square$ are open-ended, which means that states, behaviors, and even the system itself is continually changing over time (Stanley, 2019). In turn, closed-loop relations allow for iterative refinement of existing concepts and features.

Sometimes, the world has unique aspects that do not enable good observation (*M*) of *R*. In the context of modern AI systems, being able to interpret the vagueness or noisiness of training data through various feedback mechanisms provides robustness and resilience to a given artificial learning application. While these aspects are included in the Wang (2019) definition, there are over 70 formal definitions of intelligence in the scientific literature (Legg and Hutter, 2007). This allows us to define intelligence beyond the standard internal model of cognition, and include definitions that encompass second-order cybernetic observers. This provides both a limitation and an opportunity: while all current definitions of intelligence are inadequate, a physics-inspired world model based on the EGR allows for an adaptive kernel that can generalize and supervise local optimum. Our candidate world model also provides us with a model that responds to both internal and external goals. World models based on the EGR are particularly well-suited to augmenting performance for certain types of modern AI systems. Sequential processes such as time-series or LLMs are a good fit. Physics-inspired models such as diffusion or flow matching is an even better fit. The emergent properties of an EGR world model are interesting. By virtue of the *S-R* and *S-R-M* relations not being deterministic, the resulting world models are generative with respect to out-of-distribution data.

To show the broad applicability of the EGRT, and cybernetic regulation more generally, we provide several examples of physical systems with statistical regularities: temporal criticality, non-normal denoising, and alternating procedural acquisition. These examples not only provide insights into the learning process but also how physical processes embodying rare events can be adapted to and learned. Overall, the EGRT provides a mechanism for developing world models consistent with a wide variety of artificial learning application domains, from Machine Learning and Diffusion Models to Reinforcement Learning. The way in which a continuously improving map between *S* and *R* unfolds over time provides an advantage over hard-coded world models. Hopefully, our work will provide a starting point for more concrete applications.

**Supplement to A "Good" Regulator May Provide a World Model for Intelligent Systems**
Alicea et.al

## Appendix A

### Description of the EGRT

Good regulators can result from a variety of mappings between the regulator (*R*) and the system (*S*). *S* and *R* are often related as a closed-loop first-order feedback, but can achieve a specific external goal via other means, including open-loop feedforward and higher-order feedback. In general, configuration $\boldsymbol{\psi}$ is a one-to-one mapping of the system, which results in producing a desired set of outputs *G* given the total set of outputs *Z* (Figure 1). The region of interest *G* is a subset of the total system Z, and for smaller areas of G are relative to Z, the more precise mapping $\boldsymbol{\psi}$ must be. The dynamics of the system relies partially on the homeomorphic mapping between *S* and *R*. System dynamics also rely on disturbances (*D*), which provide a variational input that drives both our system (*S*, input represented by $\phi$) and our regulator (*R,* input represented by ρ) away from static equilibrium.

The example in Figure 1 only considers external perturbation. Our examples in Figure 3 feature disturbances that are hard to integrate using the EGRT formulation, which provides a partially adversarial example of good regulation. Returning to the Daisyworld example, periodic oscillations driven by internal dynamics rather than external perturbation (Nevison et.al, 1999), or recycled perturbation with the model itself, can also play a role in the output and thus affect the overall mapping between *S* and *R*. Figure 5 demonstrates how these parameters map to a good regulator model of gradient descent and Q-learning algorithms. This assumes that good regulation can be achieved through programmatic means. Indeed, there are separate steps between observation (*M*) and *S*. A good regulator model of algorithmic learning is not a special case of a control-theoretic formalization, as there is an open-ended element of the good regulator not captured by control theory.

## Appendix B

### Regulator Capacity

The capacity of a regulator can be described as the number of discrete states represented in the model (see Figure 2). The relationship between the discrete state definitions between *S* and *R* is best characterized as a first-order *S-R* closed-loop relation. If a given model can match the number of states in the system, then the model is said to have appropriate capacity to be a good regulator. But this relationship is characterized as a rough equivalency, as a model with many more states than the system under regulation can produce overspecified outputs. In terms of variety, models with fewer states than the system destroyed the full variety of the systems.

Conversely, when $R$ has a number of states greater than the system destroys $R$'s capacity to regulate. In the case of a world model, capacity must be maintained globally across all local model configurations. Therefore, world models of a system must be closely matched and ergodic: a local model that matches the states inherent in the corresponding subsystem, and a global model that matches the appropriate number of states over all subsystems.

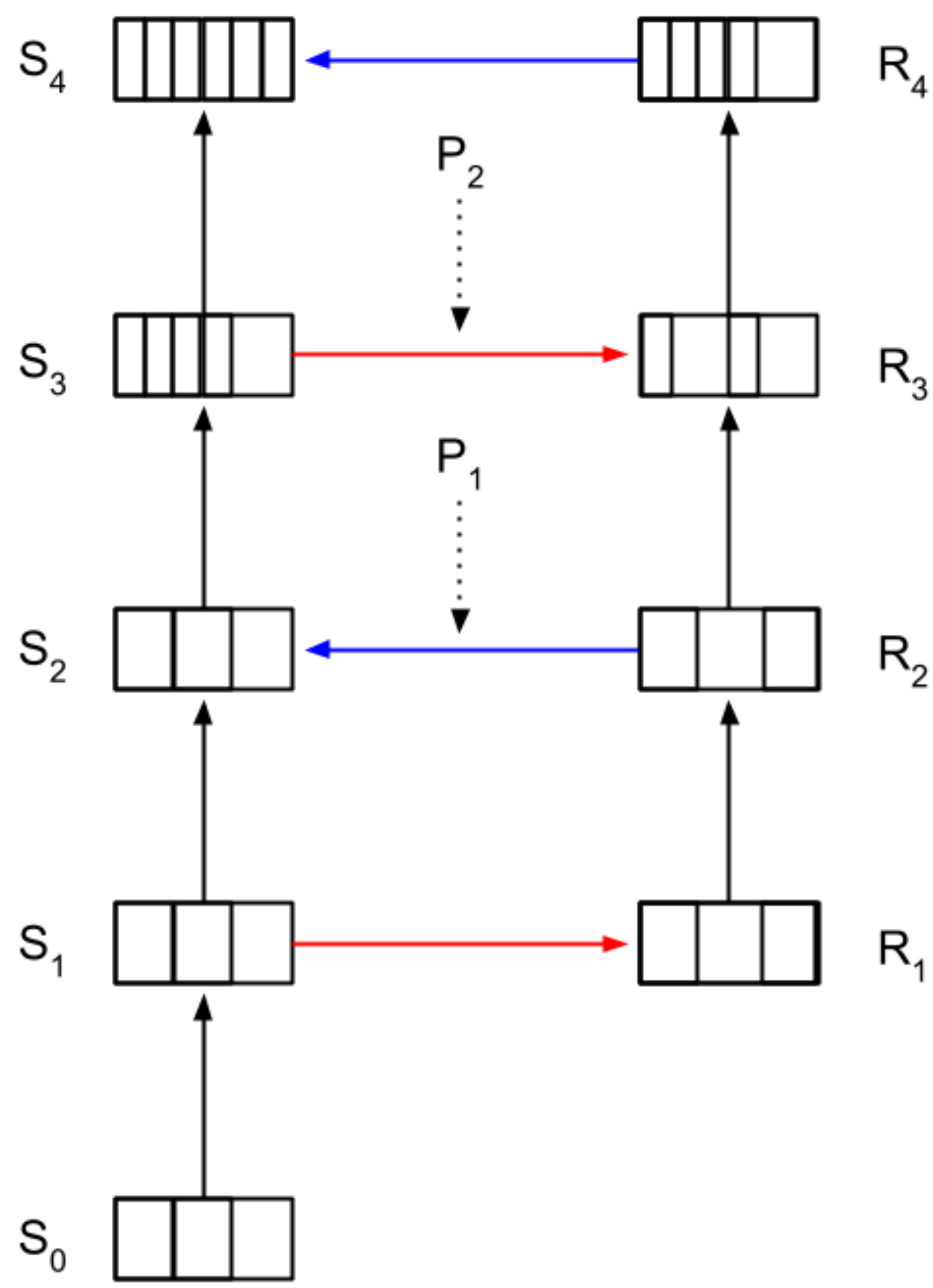

Figure S1. The acquisition of $S$ by $R$, the comparator of $S$ by $R$, and perturbation of $S$-$R$ closed-loop relation over time. $S_0$ represents the initial condition of a system. $S_n$ and $R_n$ are represented at every time interval in their interaction and their corresponding number of discrete states. State information is updated in $S$ as the system evolves, and in $R$ as information is acquired by $S$ (red arrow) and compared by $R$ (blue arrow). Perturbations can impact feedforward (acquisition) and feedback (comparator) by disrupting the quality of information, and ultimately destroying variety.

**Components of Good Regulation**

Consider closed-loop feedback in the form of a reversible bijective map (Figures 1A and 2). $M$ is the internal model, in which the observer creates a model that approximates the components of $R$. Each $S$-$R$ closed loop relation serves as a local $M$ of $S$. An ensemble of closed-loop motifs can be composed to provide a global *M* of *S*, or a world model (Figure 1B). M is also referred to as an internal model (Francis and Wonham, 1976), and can provide a state-space approximation as assumed in Figure 2, or a more specific model of information processing. Good regulation requires four additional components: input, feedback, feedforward, and output. **Input** provides information to the *S*-*R* closed-loop relation: R samples S at a variable rate and either provides a means to match the aggregate behavior of S. As the number of samples increases, states of the system are acquired by R. The ability of *R* (or alternately *M*) to approximate *S* determines aspects of operational closure, thermodynamic openness, and open-ended maintenance (Maturana and Varela, 1980; Jaeger et.al, 2024). Better representations provide more information about local thermodynamic fluctuations, identification of states, and communication with adjacent closed-loop feedback systems. **Feedback** is the interplay between *S* and *R*, which improves the quality of regulation over time. Positive feedback allows *R* to acquire the states of *S* and approximate the behavior of *S* more reliably. Negative feedback disfavors matching by obscuring the states of *S* or approximation of *S*'s behavior. Negative feedback is a lack of improvement towards internal or external goal (*G*), while positive feedback is improvement in the direction of *G*. **Feedforward** contrasts with feedback by eliminating *G* as part of the interaction between *S* and *R*. This eliminates the continual improvement and verification mechanisms between *S* and *R*, and allows for nonlinear aggregation of behaviors such as criticality that do not directly approximate *S*. **Output** is the behavioral output of *R* and alternately *M* that approximate *S*. As *R*'s (and alternately *M*'s) observations of *S* accumulate, *R*'s (and alternately *M*'s) behavioral output become more accurate. Feedback allows for a one-to-one mapping, while feedforward allows for a nonlinear, or temporarily misaligned mapping.

Systems with feedforward but not feedback do not have access to nonlinear behavioral modes nor the refinement of representational states. In Figure 2, the relationship between *R* and *S* assumes closed-loop feedback. As an open-loop feedforward system simply excludes the feedback component, *R* reproduces the behavior of *S* in a semi-blind manner. The avalanche and alternating procedural learning examples demonstrate how this works in practice: in the avalanche model, our controller is totally reliant on the observation of an avalanche, thus reducing its sensitivity to out-of-distribution events. The more rare an avalanche of a specific magnitude, the less accurate the internal model representation of *R*. In the case of alternating procedural learning, feedforward-exclusive regulation does not allow for relearning nor the resulting generalization to occur. The internal model representation will simply be destroyed during the unlearning phase, with regulation being either suboptimal or non-existent. This is one reason why we present the notion of learning-unlearning-relearning as a means to consolidate a representation across contexts. Our conceptualization of unlearning is similar to what happens in Hopfield Networks. As Recurrent Neural Networks that are able to exhibit spin glass phase transitions (Bovier and Niederhauser, 2001), Hopfield Networks undergo an unlearning phase that acts to optimize the learning process (Hopfield, 1982), serving as a function similar to sleep and dreaming in animal brains (Ellis et.al, 2021).

## Appendix C

### Sandpile Avalanches

Sandpile avalanches can be modeled using a time-series conforming to a stochastic power law. Simulating a suitable time-series requires us to generate a *1/f* power law, shown in Figures 3A-C and Figure S2 (SciLab code shown in Table S1). A stochastic power law is a challenge for the model because the system cannot be fully approximated in advance. Avalanches have no characteristic length scale, and as such cannot be easily predicted. We utilize two types of communication between models: feedforward and feedback. These represent open-loop and closed-loop acquisition from the system, respectively. One aspect of the world model is how the model is continuously updated both locally and globally.

The feedforward and feedback components of the model are shown in Figures 3B and 3C, respectively. In these graphs, the model tries to approximate the system from the previous time points. From these plots, feedforward tends to match the actual state of the system with lag, while feedback also matches the system in this manner but tends to overshoot the amplitude. Feedforward is the closest analogue to a predictive model trained on the data, while the feedback signal is less effective at representing the amplitude of the system. This is similar to approaches such as data-driven control (Pan et.al, 2022), where stochastic disturbances are introduced to *S* where *R* only represents the first two moments of the time-series. Non-ergodicity and criticality (Cherstvy and Metzler, 2014) are highly relevant to the non-Gaussian diffusion processes shown in Figure 3I-3M.

In Figure S2, we use a similar methodology used in Figure 3A-3C, but run a more formal stochastic simulation for $10^4$ moments for each component of a *S-R* closed-loop relation. We do not show the frequency spectrum, but rather tease out the components of the cybernetic interaction. This yields three graphs: a characterization of the output behavior (Figure S2A), the acquisition or mapping from S to R (Figure S2B), and the comparator or feedback from R to S (Figure S2C). In Figures S2A-S2C, the parameter *-a* (values -0.5 and -1.0) is used to characterize the amount of noise in the simulation. Different components of S-R closed loop relations allow us to consider forward and inverse criticality. Forward criticality is enabled by feedforward processes: the output of R is generated via acquisition from S, but R does not act as a comparator. Instead, there is an accumulation of output that characterizes system behavior. These sequential outputs are the observed avalanches that define criticality. When feedback is established between R and S, more specifically the comparator aspect of R, this leads to the potential prediction of output. Since our experiment assumes a highly stochastic S, the ability of R and even M to match the current state of S in an ergodic manner is limited.

How does a model know when an avalanche is most likely given a probability distribution? For Figure S3, we sort the time-series by amplitude. This provides a smoothly increasing exponential curve that provides general information about the frequency of avalanches. This represents a blind signal which averages global information about the system. The model takes this information and operates on a threshold. The threshold point is based on the probability density function, and determines when an avalanche will occur at a 100% probability. This threshold point is the most conservative estimate of when an avalanche will occur: due to the non-stationary aspect of criticality, avalanches are likely to occur before this point. Therefore, Figure S3 is to be viewed in comparison to a sandpile model time-series.

Table S1. Code for generation of sandpile avalanches and closed-loop first-order feedback components for regulating a time-series. Implemented in SciLab 2024.0.1

```
t = 1:length(time);
// define time series.

for i = 1:10000;
        S(i) = t(i)^-e;
end
 // power noise for a given exponent, fractional or a whole number < 6.0.

S = grand(1,"prm",S);
// randomizes the frequency of events in the power spectrum across time. Forward map.

plot(t,S) // plot shown in Figure 3A.

for i = 1:9999;
        PFB(i) = ((S(i) + S(i+1))/2);
end
// description of the positive feedback component of the forward map. Reverse map 1. Plot
shown in Figure 3B.

for i = 1:9999;
        NFB(i) = ((S(i) - S(i+1)));
end
NFB = abs(NFB);
// description of the negative feedback component of the forward map. Reverse map 2. Plot
shown in Figure 3C.

t = 1:10000;

for i = 1:10000;
        A(i) = t(i)^-1;
end
A = grand(1,"prm",A);
// reproduce original time-series distribution of avalanches.
```

```
for i = 1:10000;
        AA(i) = t(i)^-0.1;
end
AA = AA';
AA = gsort(AA,"c","i");
// Produces a threshold function model used to detect the probability of avalanche. Analogous to a cumulative probability distribution at a more densely distributed set of events. Shown in Figure 3D.

T = 1:10000;

for i = 1:10000;
        S(i) = t(i)^-e;
end
S = grand(1,”prm”,S);
S = matrix(S,100,100);
// smooth system observations by three orders of magnitude.
S_m = mean(S,2);
T = 1:100;
plot(T,S_m)
// comparison of directly-observed sandpile avalanche system versus the smoothed model used for avalanche prediction.
```

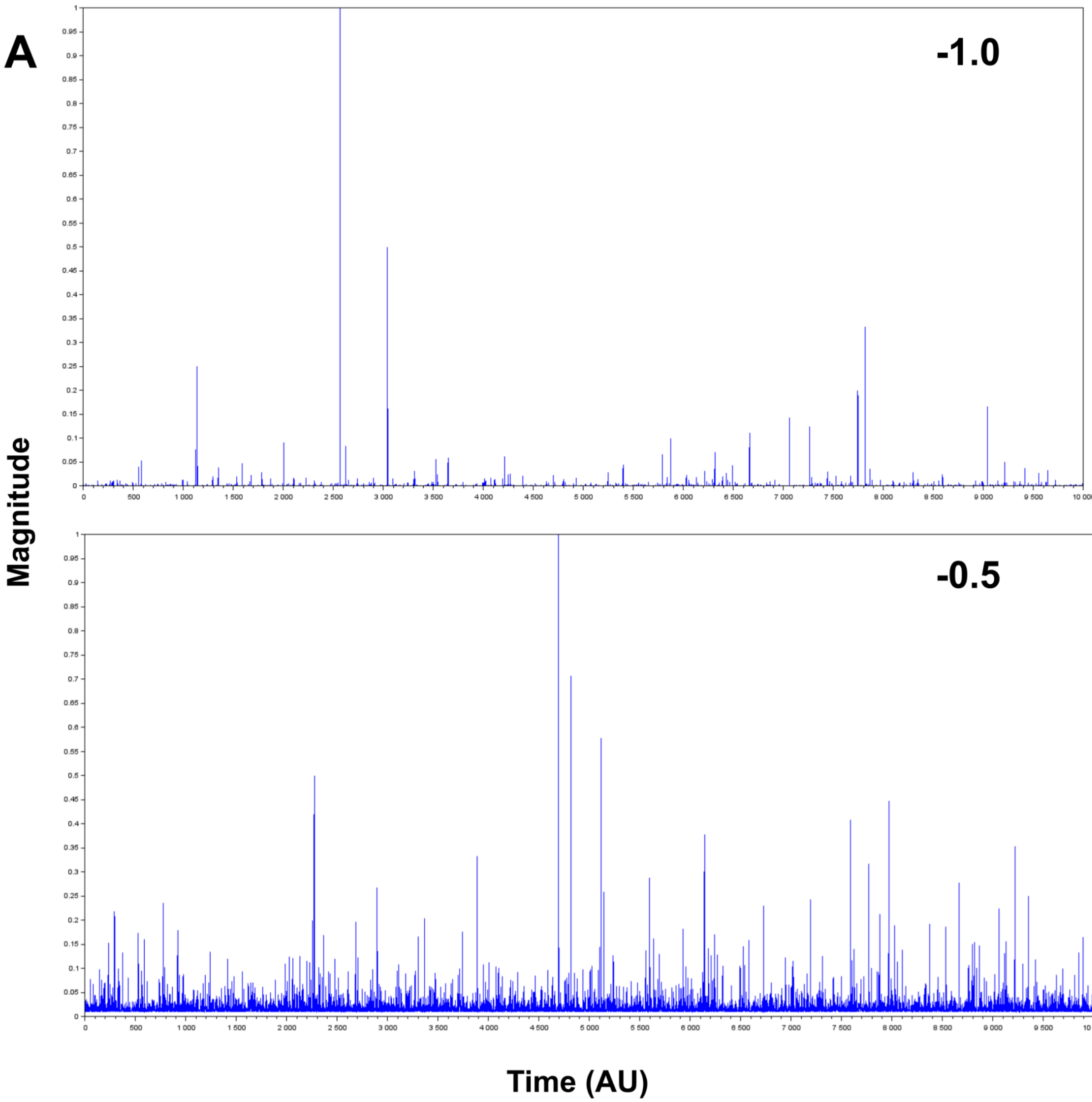
A
-1.0
-0.5
Magnitude
Time (AU)

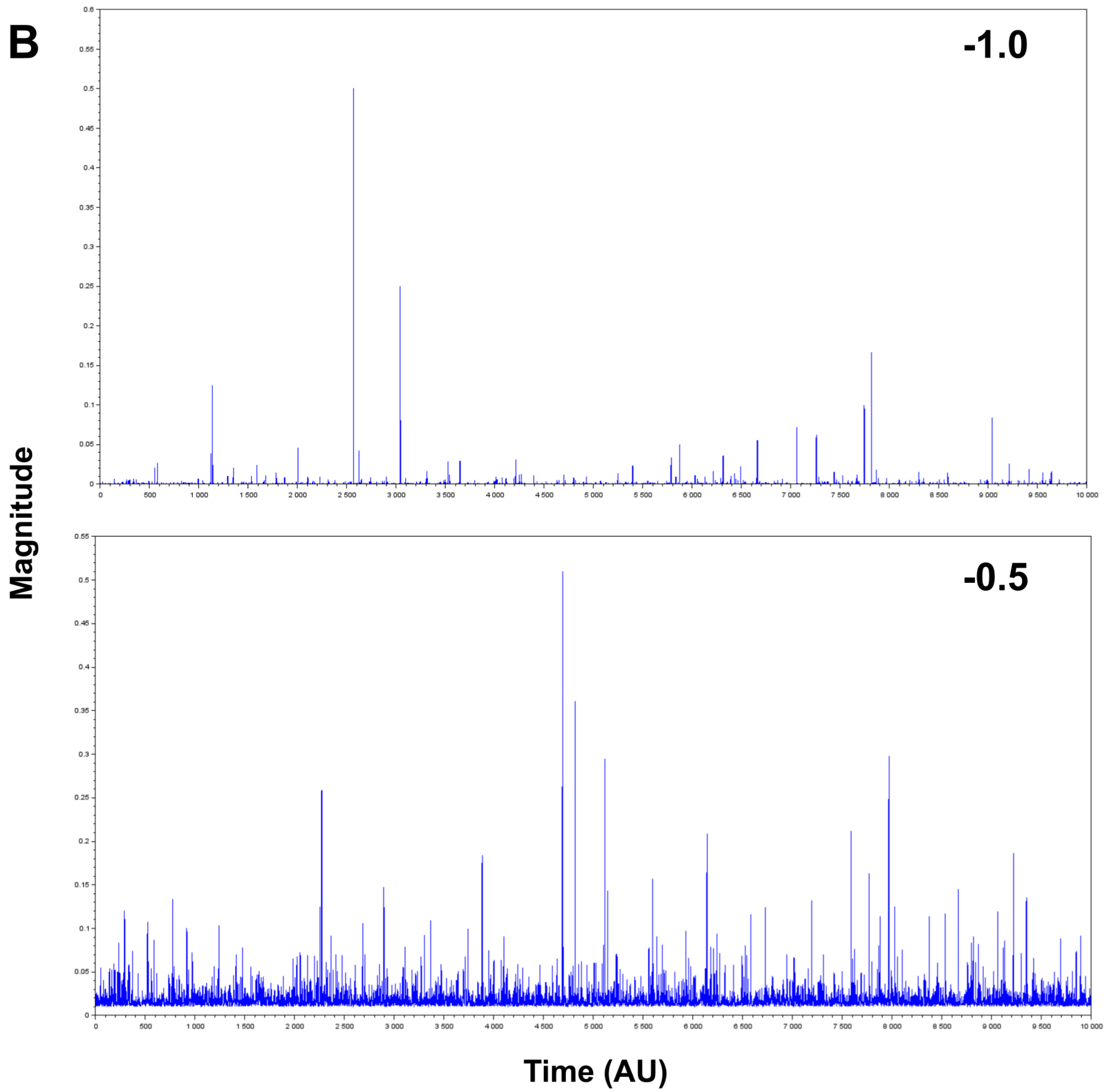
B
-1.0
-0.5
Magnitude
Time (AU)

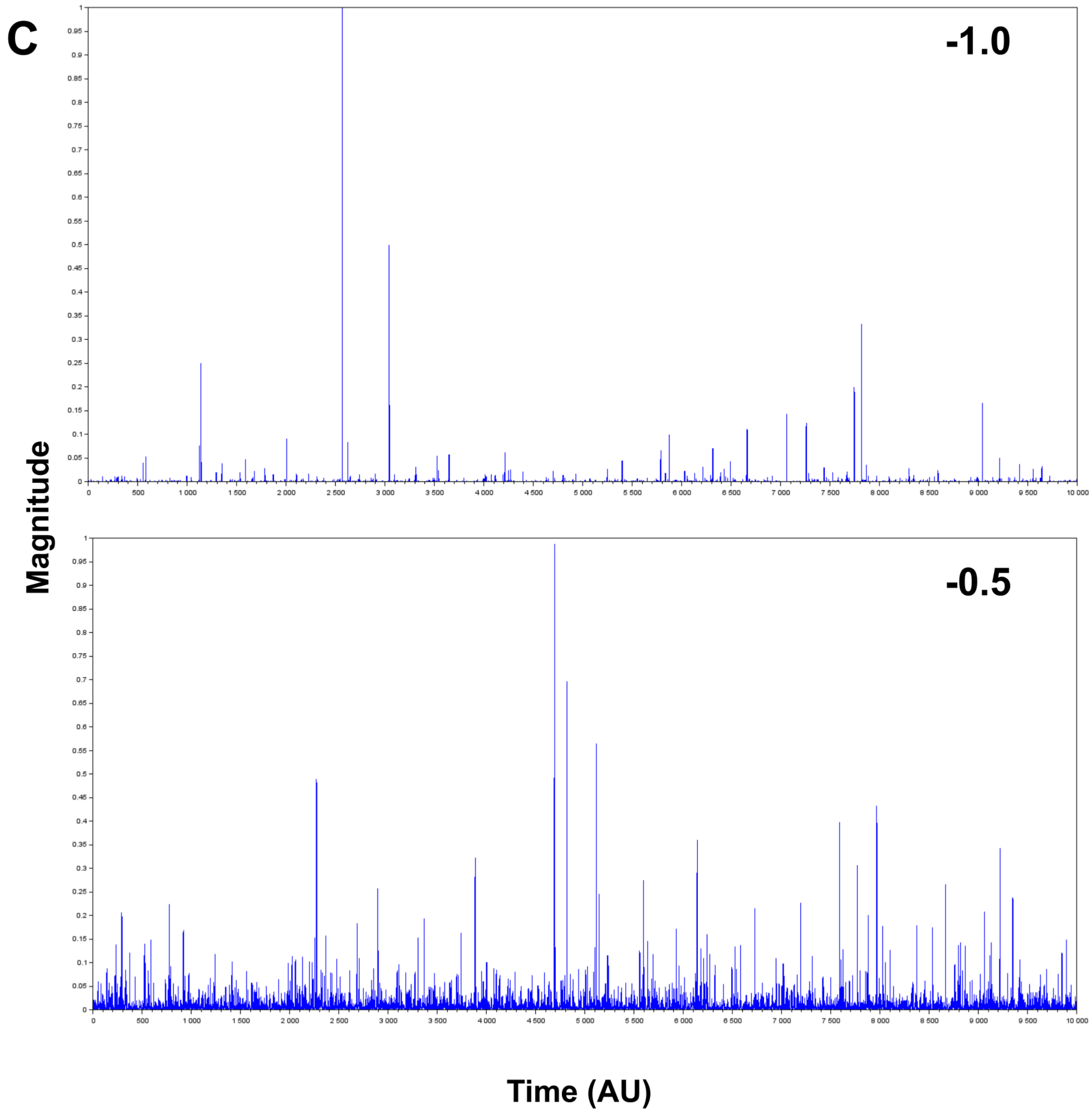

Figure S2. Plots of the sandpile model and example of the forward diffusion model at exponent values 1.0 and 0.5 (Figures S2A-S2C). A: Observation of the external behavior of a basic *S*-*R* relation. B: acquisition of *S* by *R*. C: comparator or feedforward component of *R*.

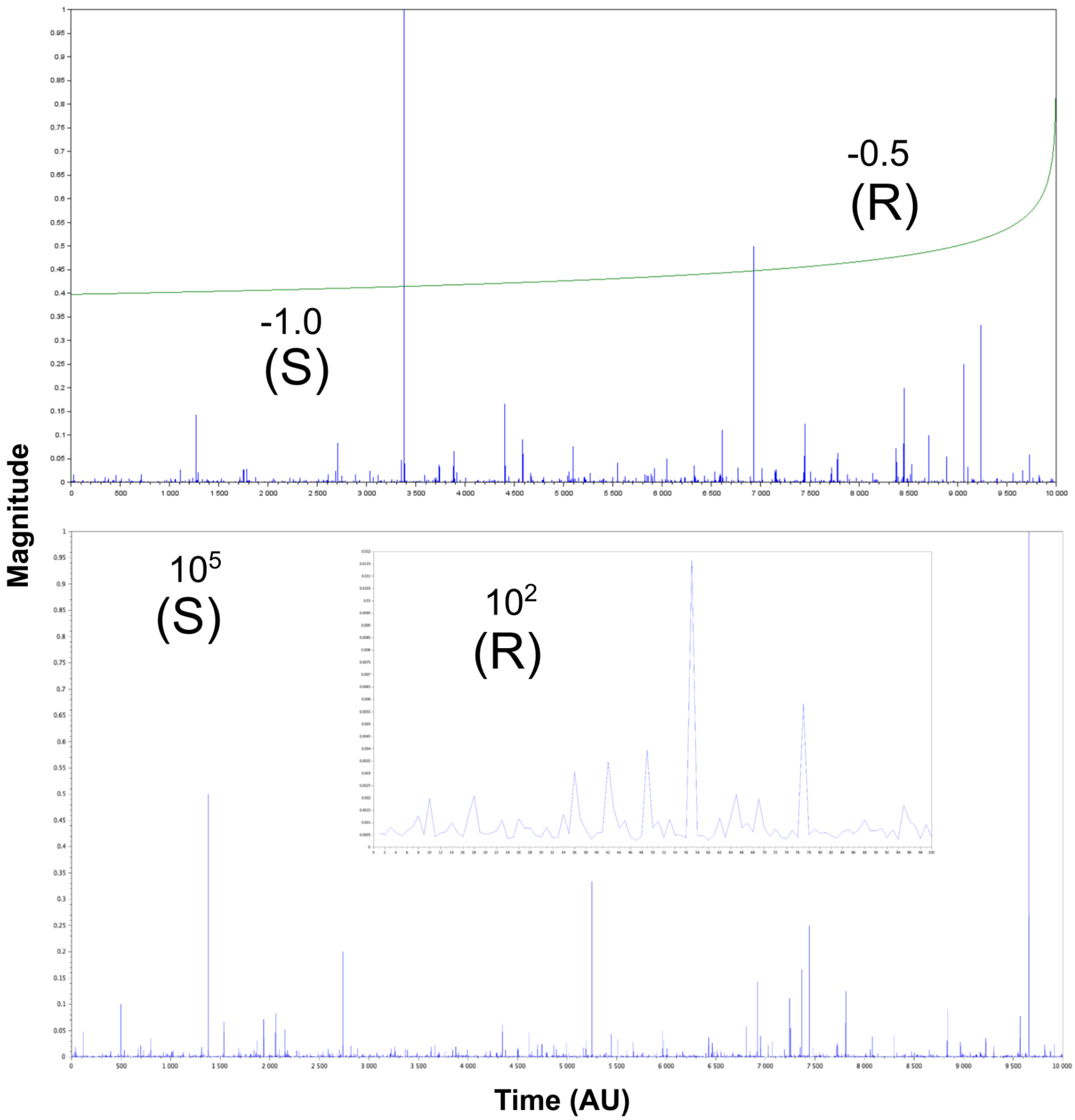

Figure S3. Two means of modeling an avalanche. Top: $10^4$ observations of the sandpile avalanche distribution (blue, system) against detection signal (green, model). Noise (-*a*) value of -0.5 is used for *S* and -1.0 is used for R Bottom: $10^4$ observations of the sandpile avalanche distribution (*S*) against $10^2$ smoothed observations extracted from the sandpile avalanche distribution (*R*, inset).

Figure 3D demonstrates two alternative representations the internal model can represent the sandpile avalanche system: as a cumulative probability distribution at multiple scales (main image) and in terms of smoothed dynamics (inset). The cumulative probability distribution represents all observed events from smallest to largest in order of probability. This is the model's observation of the system: the idea is to find the point at which a proper response to an avalanche comes with an acceptable penalty (e.g. false positive). To represent this bifurcation (when to respond and when not to respond), the model resamples the time-series and uses the linear to superlinear threshold to determine the location of this bifurcation. In Figure 3D (top), the system is observed at a power law exponent of -1.0, while the approximation is sampled at a power law exponent of -0.5. Figure 3D (bottom) shows an alternate approach, in which the system's history ($10^5$) is summarized as a 100-point smoothing process ($10^2$). The smoothed series provides a rough summary of general trends in the system's historical dynamics without requiring full observation.

A final note on the power law function that produces our probability distributions for avalanches. The default exponent for generating power law noise is 1. However, we can change this value to generate different properties of the time-series. As the exponent value approaches 0.0, the distribution features a lower maximum amplitude overall, with a distribution that is more normal. As the exponent value increases past 1.0, the distribution becomes more sparse, and for values above 5.0, an extremely sparse distribution with very few data points of high amplitude resembling black noise (Amaba et.al, 2023) results.

**Diffusion and Good Regulation**

As with good regulators, machine-learning diffusion models have connections to non-equilibrium thermodynamics. Diffusion models operate by leveraging noise for generalized learning. As a stochastic process, diffusion provides a means for predicting criticality and other hard computational problems. Our demonstration shown in Figure 3D-3M is consistent with modern diffusion models in ML. Given a training set of images, noise is progressively added until a full white noise mask for each image is achieved. The original image is then recovered, with the noise providing variety for the diffusion model. This can be stated in terms of the EGRT: *S* sends a progressively noisy signal to *R*, which is corrected through feedback from *S*. While normal statistical distributions are Gaussian, Power distributions approximate the non-normal aspects of *S*. In Figure 3, we use both Gaussian and Power noise to show the nonlinear effects of forward denoising.

Diffusion models in particular are closely tied to thermodynamic processes (Ambrogioni, 2024), which is in turn tied to the acquisition of *S* by *R*. Acquisition is driven by a given probability distribution and provides a means to acquire increasing more state information from *S* (Alonso et.al, 2024). In the case of machine learning diffusion models, random components of the probability distribution, acquired from making the training set and resulting embedding intentionally more variable, are necessary for arriving at the global minima. Our forward denoising scheme is inspired by the Tiny Diffusion toolbox (www.github.com/tanelp/tiny-diffusion), which is a PyTorch implementation of probabilistic diffusion models for 2D datasets. Even though diffusion models typically diffuse images around their salient features, we dissolve the entire image into noise, which is optimal for demonstrative purposes. While our demonstrations show how an image can be embedded in a noised latent space, recovery is by no means without good regulation. Interpolation during image reconstruction, smooth transition between nearby data modes during reconstruction, generate artifacts not present in the original data. This results in out-of-distribution behavior that can lead to effects such as hallucinations (Aithal et.al, 2024).

The first step in diffusion models is the forward noising process (Wang et.al, 2024) shown in Figure 3E-3N. Forward sampling involves introducing increasing amounts of noise into an image. This obscured image is then used to build a latent space, or model of variation for the training set. Recovery can proceed through techniques such as variational inference (Ho et.al, 2020). In Figures 3E-3I, the progressive introduction of noise creates more variety in the feature space: in EGRT terms, variety is created in the model so that variety can be destroyed (or overcome) in regulating the system of training images. This is one reason why diffusion models outperform other types of neural network (Wang et.al, 2024). Noising can be approximated by Fourier models, which decomposes this variety into frequency components (Tancik et.al, 2020).

To model a forward noising process and demonstrate the type of probability distribution produced by this process, we utilize two methods for noising a training image. Figure 3E-3I features a grayscale image of W. Ross Ashby masked with a Gaussian signal using the *random.rand()* method in Python. In this case, the Gaussian noise is uniform, but is blended using the alpha parameter in NumPy. The noising process increases over the series of images, adding to the variation of the model. Figures 3J-3N uses power spectral noise characterized with shape parameter a to model non-linear noising effects. This is analogous to looking at the effects of criticality on the avalanche model shown in Figure 3A-3D. In the power noise example, the noising process is ordered according to decreasing values of a (3J = 0.8, 3K = 0.6, 3L = 0.4, 3M = 0.2, 3N = 0.01). Power noise is generated using the *random.power()* method in NumPy (Table S2). In the case of power noise, variation is introduced to image features in different ways (Vafaii et.al, 2024), creating variation within variation, and more accurately characterizing the system.

Table S2. Power law noise with an exponent of 1.0. Implemented in Python 3.7 and TensorFlow.

```
import numpy as np
import random
import matplotlib.pyplot as plt
power_noise = np.random.power(1,113100) //exponent value, number of pixels.
a = np.reshape(power_noise, (377, 300)) // matches size of image.
print(a)
plt.imshow(a, cmap='gray')
plt.show()
```

```
pip import tensorflow as tf
from tensorflow.keras.preprocessing.image import load_img
wrossashby = load_img('w-ross-ashby.png')
display(wrossashby)
plt.imshow(wrossashby, cmap = 'gray')
plt.imshow(a, cmap='gray', alpha=0.75)
```

**Appendix D**

**Alternating Procedural Learning with Force Fields and CMYK Gradients**

One example of alternating procedural learning can be demonstrated through the acquisition of changing force fields. Similar to the avalanche model, the system exhibits multiple modes over time, where the shift to a new force field is both stochastic and novel to the model. Unlike with diffusion models, we cannot simply add variation to the training. In this case, we need to introduce new phases to the model, which allows for a world model of all multiple configurations of a generic phenomena. In the case of alternating procedural learning, we need both a model and an embodied morphology. This allows our model to be oriented to conditions in the system in a particular manner. In this way, generalized learning can be achieved by introducing new regimes over time (see Figure 4, top). In terms of the *S*-*R* closed-loop relation, *S* presents different orientations and distances to an internal or external goal, while *R* tries to match and respond to this variety using a form of differential learning. Once consolidated into long-term memory, motor behavior characterized by *R* becomes tightly matched to the mechanical world characterized as *S* (Mussa-Ivaldi and Bizzi, 2000). Returning to the Daisyworld example, alternating procedural learning might allow domain $\boldsymbol{\psi}$ to acquire a wider range of stable operating states.

Alternating procedural learning can also be used to identify CYMK gradients. As with the force field example, *R* is embodied in a morphology that is exposed to *S* in multiple ways. Transitions between alternating regimes are discontinuous, and are not easily predicted by *R* without reference to all possible states of *S*. Exploration of the CYMK gradient resembles a Braitenberg Vehicle (Braitenberg, 1984) in its ability to follow sensory gradients using a simple input-output relation. In this case, the comparator function of *R* in the *S*-*R* closed-loop relation can occur both inside and outside the embodied system. Inside the vehicle, cross-talk between different sensors introduces a comparator function in a feedforward configuration. An external comparator comes in the form of a sensory gradient *S* and the effector output of the vehicle *R*. The internal mapping between *S* and *R* acts as an internal model *M*. This allows a regulatory domain to move an agent towards *G* on a highly deterministic *S*.

**Appendix E**

Table S3. A list of terms and definitions.

| Term | Definition |
|---|---|
| System (*S*) | A system (*S*) is a collection of attributes that represent a world. In the EGR, *S* contains a variable number of discrete states over time. |
| Regulator (*R*) | A regulator (*R*) is an approximation of a system (*S*) that acquires state information from *S*, and verifies the content of *S* via feedback. *R* is an open set that must approximate the true number of states in *S*. |
| Internal Model *(M*) | An approximation of *S* via *R*. Enables control of *S* by providing a supervisory (purposeful) component to *R*. *M* approximates time-averaged information for *S*, tries to approximate ergodicity. |
| Global Domain (*Z*) | A global domain which consists of all possible states of *S*. The objective of an *S*-*R* relation is to restrict Z to a smaller region without destroying the variety of *S*. |
| Restricted Domain (*G*) | The objective of a regulator, or restriction of domain *Z* by *R*. G is contained as a subset of *Z* without excluding significant information about the global state of *S*. |
| Variety | A sufficient number of discrete states relative to the system *S*. Sufficient variety involves *R* being able to match the states of *S*. |
| Good regulation | In general, a closed-loop relation between R and S. Good regulation can also occur via an open-loop relation, or include the role of an internal model (observer, defined as *M*). |
| Control | For the EGR, control proceeds by *R* finding a solution *G* on *Z*, or *M* supervising *R* to produce a controlled output. |
| Ergodicity | Where the time-averaged behavior of all states in *S* are equiprobable. This leads to allostasis, or systems with multiple equilibrium points. |
| World Model | A means to learn and represent the global environment in a Machine Learning (ML) or Reinforcement Learning (RL) system. World models are predictive in that they allow an ML/RL agent to simulate and understand the context of its behavior. |